# Highly accurate first-principles models of molecular interactions

Alexander G. Donchev,[1,†] Elizabeth Decolvenaere,[1] Michael Bergdorf,[1] Xinyi Guo,[1] Rian C. Kormos,[1] Ka-Hei Law,[1] Amrut Nadgir,[1] Andrei Osipov,[1] Andrew G. Taube,[1] John L. Klepeis,[1] Michael P. Eastwood,[1] and David E. Shaw[1,2,†]

[1] D. E. Shaw Research, New York, NY 10036, USA.

[2] Department of Biochemistry and Molecular Biophysics, Columbia University, New York, NY 10032, USA.

† To whom correspondence should be addressed.

David E. Shaw

E-mail: David.Shaw@DEShawResearch.com

Telephone: (212) 478-0260

Fax: (212) 845-1286

Alexander G. Donchev

E-mail: Alexander.Donchev@DEShawResearch.com

Telephone: (212) 403-8298

Fax: (646) 873-2298

## Abstract

It is a longstanding challenge to develop first-principles models—interaction potentials derived only from quantum mechanical data—that can be used in molecular simulations to reliably predict condensed-phase properties to within experimental error. Here, by fitting to state-of-the-art quantum mechanical data extremely accurately, we develop first-principles models for eleven organic molecules and water, and determine their liquid-phase thermodynamic properties. Notably, close agreement with experiment is observed not only for pure liquids and mixtures but also for differences between isotopologues (including light and heavy water), which arise from nuclear quantum effects. We also find that the inclusion of an explicit many-body dispersion term in the models is critical for accurately predicting the properties of organic liquids, and at room temperature is substantially more important than nuclear quantum effects. Our results have substantial implications for future development of first-principles models, including machine-learned potentials.

## Introduction

Molecular dynamics (MD) simulation has become an invaluable tool for understanding the atomic-scale behavior of diverse condensed-phase systems across biology and chemistry. From fundamental investigations of protein folding[1] to those of liquid-liquid critical points in water,[2] the spatial and temporal resolution of MD simulations has led to insights that can be difficult, or impossible, to obtain through experimental methods alone. In addition, MD simulations are now routinely used in industry, for example to evaluate the strength of protein-ligand interactions in drug discovery.[3] The accuracy of MD simulations, however, depends crucially on the energy function, known as a force field, that dictates how the interactions between the molecules are modeled,[4] and currently used force fields are widely acknowledged to have serious limitations. Development of force fields that can predict properties of condensed-phase systems much more accurately—roughly speaking, as well as or better than experimental approaches—is a longstanding goal with real potential to impact human health.

In principle, it might appear straightforward to achieve this goal by using quantum mechanical (QM) calculations to generate reference data,[5–8] such as molecular interaction energies, and then fitting force fields to this data, but the practical challenges have proved to be enormous.[9] To make their fitting and evaluation computationally efficient, most commonly used force fields have fairly simple functional forms that omit important physical effects; they typically assume, for example, that a molecule's electron distribution does not depend on its surroundings.[10] These force fields do not have the capacity to simultaneously fit all of their training data with high accuracy, so some compromises must be made. The typical approach is to add experimental condensed-phase data to the training set and to weight this data in the fitting so that the force field reproduces it fairly well, in effect by engineering a favorable cancellation of the errors that arise in modeling the underlying molecular interactions.[4,11] This approach has produced useful force fields, but has well-known limitations due to the lack of experimental data for many molecules, and even for those where data is abundant (like water), the commonly used functional forms do not have the capacity to fit all of it well simultaneously.[12]

To avoid these limitations, it has long been hoped that models with high-capacity functional forms could be trained exclusively on QM data to describe the energetics of molecular interactions sufficiently accurately that the resulting first-principles force fields (i.e., force fields fit to QM data only) would give excellent results in condensed-phase simulations, without having to rely on substantial cancellation of errors.[13,14] Although recent work has pointed to the importance of accounting for certain physical effects that are omitted from most force fields, such as many-body dispersion (MBD; dispersion beyond the pairwise-additive level)[15–17] and nuclear quantum effects (NQEs),[18–21] the magnitude of these effects, and whether other omitted effects must be modeled, have been unclear, as has the specific nature of the functional forms required and the QM training data needed for fitting.

Here, for a chemically diverse set of neutral small molecules (Fig. 1A), we present a new first-principles force field (termed *DES-XFF*) fit with extremely high accuracy to state-of-the-art QM data, largely computed at the coupled-cluster level. The DES-XFF models share a common functional form and fitting protocol, but their parameters are optimized separately for each pure liquid and each binary mixture studied. DES-XFF contains an explicit energy term to represent the MBD interactions, and NQEs are included by applying path-integral MD (PIMD).[22,23] Predictions of condensed-phase properties by DES-XFF, including free energies of solvation, are in excellent agreement with experiment for both pure liquids and liquid mixtures. Notably, our models advance the state of the art by also reproducing experimentally measured differences in the thermodynamic properties of light and heavy water, as well as the molar-volume isotope effect in several organic liquids, which arise from NQEs. Passing these stringent tests not only shows that the DES-XFF models accurately account for NQEs, but also provides evidence that the accuracy of DES-XFF in the condensed phase does not rely on a cancellation of errors between NQEs and other physical effects, such as MBD. This suggests that, although additional effects could be incorporated into DES-XFF, its current level of complexity is likely to be sufficient to obtain accurate condensed-phase results for other small neutral organic molecules. It also means that DES-XFF can be used to quantify the importance of different physical effects in determining condensed-phase properties. We find in particular that MBD effects have a substantial impact on the condensed-phase properties of most liquids (more so than NQEs), strongly suggesting that their inclusion will be important in other first-principles models, including those that are machine-learned.[24]

## Development of a first-principles force field

Development of an accurate first-principles force field requires not only a high-capacity functional form, but also a training set that is sufficiently large and accurate to capture the relevant interactions, yet still computationally affordable to generate. The DES-XFF models used for the condensed-phase results presented below were parameterized independently for each pure liquid and each binary mixture, allowing the parameters of a given molecule to differ between its neat liquid and mixtures. Training used a state-of-the-art QM dataset (see sections IX and X of the Supplementary Information (SI) for details on the methods used to generate the dataset) that, in addition to energies of monomer and dimer configurations computed at the coupled-cluster level, includes energies of multimer configurations consisting of clusters of up to 40 molecules. The multimer energies are the sum of a many-body (nonadditive) component computed with second-order Møller–Plesset perturbation theory (MP2)[25] and a two-body component computed at the coupled-cluster level, and provide crucial information about the nonadditive component of intermolecular interactions.

DES-XFF uses a functional form (see sections I–IV of the SI; the fitting procedure is described in section V) that enables it to fit the training data to extremely high accuracy (see Fig. 1, panels B–D for the distribution of errors when comparing DES-XFF to QM). Atom-centered diffuse multipoles of up to third order (octupoles)[26] allow for the reproduction of both the anisotropy of interatomic interactions and the charge-penetration effect that arises when electron clouds overlap.[26,27] An important feature of our approach is that the multipole parameters are not static, and instead depend on the positions of nearby atoms—both those within the same molecule and those belonging to neighboring molecules. These geometry-dependent multipoles provide capacity for modeling many-body effects, and are complemented by an explicit polarization model, based on induced diffuse dipoles, for handling induction effects. The geometry dependence of the multipoles adds substantial capacity to our DES-XFF models, allowing local variations in the atomic environment to influence longer-range interactions, similar in spirit to message-passing methods in some graph-based machine-learned potentials.[28–30] Intramolecular energies are modeled as a fourth-order multivariate polynomial in terms of internal coordinates corresponding to the bond lengths, valence angles, and dihedral angles.

An important feature of DES-XFF is that it treats MBD explicitly, using a damped version of the Axilrod–Teller–Muto three-body potential,[31,32] which we refer to as DATM. Only triplets of heavy atoms contribute to this term (see section IV.G of the SI). The DATM parameters were trained on MBD energies of periodic liquid systems predicted by a coupled fluctuating dipole model (CFDM).[33,34] Although DATM has a three-body functional form, it thus effectively accounts for dispersion contributions beyond the three-body level. The CFDM model is itself parameterized to accurately reproduce QM polarizabilities and long-range three-body dispersion coefficients calculated using time-dependent density functional theory (DFT). Although there have been some efforts to add explicit MBD terms to DFT,[35] in crystal structure prediction studies,[36] and in models for a few pure liquids,[15,16,37] an explicit MBD term has not typically been included in first-principles force fields.

A key expectation for DES-XFF, or any first-principles model, is that it does not rely on a single fortuitously chosen set of force field parameters, but rather that predictions for condensed-phase properties made with different models are similarly accurate, provided these models fit the underlying QM data with similar accuracy. Using our nonlinear fitting procedure with multiple independent initializations, we generated an ensemble of five DES-XFF parameterizations for each system—each neat liquid and each binary system was parameterized separately—by selecting the models with the lowest loss on the QM validation set (see section V.A of the SI). We found that all five models, despite having different model parameters, reproduced the QM data with similar accuracy. The valence and MBD parameters were held fixed across the five members of each ensemble (see section V of the SI). The condensed-phase results presented below are either for the model having the best fit to the QM data for a given system or for the full set of five models; condensed-phase results across the models are mostly indistinguishable.

Although DES-XFF is substantially more computationally demanding than conventional molecular mechanics force fields, our GPU implementation makes fairly long condensed-phase simulations practical. On a single NVIDIA RTX PRO 6000 Blackwell GPU, for example, classical MD and PIMD simulations of a periodic box of liquid water with a side length of 30 Å achieve estimated throughputs of 48 and 16 ns/day, respectively (see Fig. VI.1 in the SI for simulation throughput as a function of system size). Even with less powerful GPUs, like the A100s used for most of the simulations reported here, it was feasible to generate trajectories sufficiently long to resolve small differences in thermodynamic properties between isotopologues, obtain precise estimates of the often small excess enthalpies and excess volumes of liquid mixtures, and quantify the relatively modest contributions of NQEs to solvation free energies.

## NQEs in liquid water

Despite water being among the most studied small molecules, it has been a challenge in MD simulation to reproduce the differences in basic thermodynamic properties between ordinary (light) and heavy liquid water that arise from NQEs. Path-integral simulations with empirical water models have captured some of these isotope effects. One study[38] with the flexible q-TIP4P/F model reproduced the direction of the melting-point shift of ice Ih upon deuteration, and another[39] with a rigid model also reproduced the direction of the shifts in both the melting point and the temperature of maximum density (TMD), and also found that assigning $D_2O$ a shorter bond length improved agreement with experiment. A particularly stringent test of a first-principles force field is to reproduce isotope effects using a single potential energy function, with isotopologues differing only in their nuclear masses. Notably, and in contrast to the typical isotope effect in organic liquids, $H_2O$ has a higher number density than $D_2O$ throughout the −30 to 100 °C temperature range studied here. It is in principle possible to independently simulate the heavy and light isotopologues of water using PIMD, and compare the predictions directly to experimental results. Due to the amount of computation this requires, however, almost all studies to date have reported the difference between predictions from PIMD and classical simulations of $H_2O$; this measure of the importance of NQEs does not have a direct experimental equivalent, but is expected to be similar to (albeit of larger magnitude than) the differences between $H_2O$ and $D_2O$. The majority of water models predict that NQEs either have near-zero impact on[40] or even reduce water density,[18,38,41–46] in contrast to what the experimental isotope effect suggests.

As illustrated in Fig. 2, DES-XFF accurately predicted the densities of both water isotopologues across a wide temperature range, from the supercooled region to the boiling point. Notably, DES-XFF predicted that the temperature of maximum density (TMD) for $D_2O$ is higher than that of $H_2O$ by 11 ± 4 °C, in agreement with the experimental difference of 7 °C. In addition, it predicted that the isotope effect on number density $\delta n = (n_{H2O}/n_{D2O} - 1) \times 100\%$ is positive across the entire temperature range, also in agreement with experimental data. At room temperature (25 °C), for example, DES-XFF predicted $\delta n = 0.5\%$ (close to the experimental value of 0.35%), whereas early pioneering work reported a value for the TIP5P model of the opposite sign ($\delta n = -1.1\%$),[47] and two other studies[44,48] found $\delta n \approx 0$ within the statistical uncertainties of their PIMD simulations. We are aware of one study (using a machine-learning model trained on the SCAN DFT functional) that predicts a positive $\delta n$ of similar magnitude to that of DES-XFF, but this method predicts a water density about 6%–7% above the experimental value.[49]

To further quantify the effect of NQEs using DES-XFF, we employed the more typical approach of comparing results from PIMD with those from classical simulations. For $H_2O$, DES-XFF predicted that the inclusion of NQEs decreases the TMD by 16 ± 4 °C, a decrease ~50% larger than that between PIMD simulations of $D_2O$ and $H_2O$. Furthermore, DES-XFF predicted that NQEs always increase the density of water, ρ, and that, as expected, this effect decreases with increasing temperature, from $\delta\rho = (\rho_{PIMD}/\rho_{classical} - 1) \times 100\% = 5\%$ at −30 °C in the supercooled region to 1.4% at room temperature to 0.9% at the boiling point. This result, which as expected has the same sign as, and is larger than, the difference in number density between $D_2O$ and $H_2O$, is in contrast to the results obtained with other advanced water models that, at room temperature, predict negative values ranging from $\delta\rho = -0.3\%$ to −3%.[18,41–46] A recent model[50] predicts a small positive value of $\delta\rho = 0.4\%$, but essentially no TMD shift. We are aware of only two models that predict a positive δρ[51,52] of similar magnitude to that of DES-XFF, but one of these models[51] predicts a water density about 4% below the experimental value, and both predict other properties to be affected by NQEs in a way that runs counter to expectations based on experimentally measured isotope effects, such as a TMD shift of opposite sign[51] or an unexpectedly small vaporization enthalpy ($\Delta H_{vap}$) shift (0.1 kcal/mol at room temperature[52]). DES-XFF, in contrast (see Table VII.I in the SI), predicted a 0.64 kcal/mol decrease in the $\Delta H_{vap}$ of water due to NQEs, and a value of 0.40 kcal/mol for the isotope effect, defined as $\Delta H_{vap}(D_2O) - \Delta H_{vap}(H_2O)$, the latter being in close agreement with the experimental value of 0.33 kcal/mol (all values at room temperature).[53] We note, however, that the absolute values of $\Delta H_{vap}$ were about 0.3 kcal/mol below experiment for both isotopologues, a point we return to below.

## Properties of organic liquids

Accurately predicting properties of organic liquids is another longstanding goal for first-principles force fields, and constitutes another important test for DES-XFF. Fig. 3 illustrates DES-XFF predictions for density and $\Delta H_{vap}$ of eleven pure organic liquids and water at various temperatures between the melting and boiling points of each liquid. DES-XFF models provided highly accurate predictions; compared to the experimental values (derived from standard databases, as described in section VIII of the SI), the mean absolute errors (MAEs) were 0.57% for the density and 0.16 kcal/mol for $\Delta H_{vap}$ for PIMD simulations using the models with the lowest QM error. (The largest density error is for acetone at 329 K (−1.9%) and the largest $\Delta H_{vap}$ error is for N-methylacetamide (NMA) at 478 K (+0.47 kcal/mol).) For comparison, an extensive benchmark of general-purpose empirical force fields[54] reported RMS errors ranging from 2% to 4% for liquid density and from 1.1 to 2.5 kcal/mol for $\Delta H_{vap}$ (Table 2 of Ref. 54). These errors are substantially larger than the corresponding MAEs obtained with DES-XFF, but the benchmark encompassed a much larger and more chemically diverse set of molecules (up to 235), some of which may present greater challenges for force field development than those studied here.

It is worth noting that for a number of small molecules, the experimental values, especially $\Delta H_{vap}$, are associated with large uncertainties that exceed the discrepancy between DES-XFF and experiment (see Table VIII.I of the SI). In those cases, it is thus possible that the DES-XFF predictions are more accurate than the experimental values. For NMA at 410 K, for example, the DES-XFF prediction of $\Delta H_{vap}$ = 13.0 kcal/mol differs from the experimental value of 12.6 kcal/mol[55] used in Fig. 3, but agrees with a more recent estimate (obtained from reanalysis of multiple experimental studies) of 13.0 ± 0.1 kcal/mol.[56] As noted above, we assessed condensed-phase performance across an ensemble of five DES-XFF parameterizations for each of the twelve neat liquids and found that the different parameterizations were similarly accurate in the condensed phase (Fig. 3, A and D); the consistency within each ensemble of DES-XFF models is such that in most cases the differences between the individual models are hardly visible in the plot.

For their ultimate applicability to biological systems, it is important that first-principles force fields can predict condensed-phase properties for mixtures. We thus used DES-XFF to predict the excess enthalpy and excess volume of seven binary mixtures (see Fig. 3, H and I for the results for an acetone/water mixture and Fig. VII.2 in the SI for all mixtures studied). DES-XFF accurately predicted the excess enthalpy as a function of composition for most of these mixtures, including the complex sign-changing behavior of acetone/water, although the excess enthalpies of methanol/benzene and pyrrole/benzene were overestimated by up to approximately 0.1 kcal/mol; no experimental excess enthalpy values were available for NMA/benzene. For excess volume, the DES-XFF results were in good agreement with experiment, except for the methanol/benzene and pyrrole/benzene mixtures, for which DES-XFF predictions were too positive by up to 0.4 and 1.1 $cm^3$/mol, respectively (see Fig. VII.2 in the SI).

## Quantifying NQE and MBD contributions to condensed-phase properties

The ability of DES-XFF to accurately predict condensed-phase properties, despite being fit only to QM data, suggests that it correctly captures the contributions of underlying physical effects such as MBD and NQEs, providing an opportunity to quantify the importance of these effects. The agreement discussed above between DES-XFF predictions and experimental results for different isotopologues of water strongly supports this notion for NQEs. Independent support for the accuracy of our treatment of MBD comes from a comparison of MBD contributions to nonadditive microsolvation energies computed in two ways: first using CFDM and second using post-MP2 coupled-cluster corrections. For microsolvation clusters of 11 ligands in water, containing an average of 214 atoms and up to 256 atoms, the RMS deviation of the CFDM predictions from the coupled-cluster corrections was 0.29 kcal/mol, corresponding to only 14% of the RMS magnitude of the post-MP2 coupled-cluster corrections themselves (2.1 kcal/mol). Further support comes from the substantially better agreement with experimental condensed-phase properties when the explicit MBD term is included, as discussed below. Although these comparisons do not rule out some cancellation of errors, they provide complementary evidence that our treatment of MBD is reasonably accurate. In this section, we quantify the impact of both NQEs and MBD on the condensed-phase properties of water and organic liquids.

In the DES-XFF predictions shown in Fig. 3B, for all liquids except water, the difference in density due to NQEs is negative (the PIMD density is lower than the classical MD density) and modest, ranging from −3.5% for ethane at 90 K to −0.1% for methanol at 338 K, with the effect decaying approximately as the inverse square of the temperature. The average difference in density due to NQEs across all of the studied liquids (excluding water) and temperatures was only −0.9% (with an even smaller value of −0.6% in the biologically relevant temperature range, defined here as 273 to 373 K; see Table VII.II of the SI). According to our results, the impact of NQEs is exceeded by that of MBD, with an average difference in density due to MBD across all of the studied liquids, including water, of −4.6% (with values ranging from −6.5% for benzene at 353 K to −3.0% for acetone at 178 K). In contrast to the difference due to NQEs, the magnitude of the difference due to MBD increased with temperature (except for water). A recent study,[17] in which a D4-parameterized Axilrod–Teller–Muto three-body dispersion term was added to a machine-learned potential, found even larger effects of three-body dispersion on condensed-phase properties, reporting average reductions of approximately 10% in liquid density and 15% in $\Delta H_{vap}$ across 39 organic liquids. It is worth noting, however, that, based on recently published coupled-cluster benchmark data,[57] the D4-parameterized three-body dispersion term (with the default Chai–Head-Gordon damping) overestimates the total three-body dispersion contributions to the lattice energies of crystalline benzene, carbon dioxide, and triazine by 34%, 59%, and 48%, respectively, if the differences between the CCSD(T) and MP2 nonadditive three-body interaction energies are taken as estimates of the corresponding three-body dispersion contributions. These findings suggest that the larger effects reported in Ref. 17 may be partly attributable to an overestimation of three-body dispersion by the D4-based model.

For $\Delta H_{vap}$, the average difference DES-XFF predicted from NQEs (excluding water) was −0.1 kcal/mol (Fig. 3E), with values ranging from −0.6 kcal/mol (methanol at 176 K) to +0.1 kcal/mol (benzene at 353 K and phenol at 434 K). At biologically relevant temperatures, the largest-magnitude difference due to NQEs (again excluding water) was −0.3 kcal/mol (imidazole at 362 K). Differences in $\Delta H_{vap}$ attributable to MBD almost always exceeded in magnitude the differences attributable to NQEs, and were always negative, ranging from −1.0 kcal/mol (benzene at 279 K) to −0.2 kcal/mol (water at 273 K), with an average value of −0.6 kcal/mol. Only for water and low-temperature methanol did NQEs play a somewhat larger role than MBD. For a given liquid, the difference due to MBD was nearly constant as a function of temperature, whereas the difference due to NQEs, when large, quickly diminished with increasing temperature. Interestingly, DES-XFF predicted that for 7 of the 12 liquids studied here, the difference in $\Delta H_{vap}$ due to NQEs is positive as the temperature approaches the boiling point (for benzene it is positive at all temperatures studied). This finding can be rationalized from the viewpoint of the competition between intermolecular and intramolecular NQEs, both of which can be related to zero-point vibrational motion. In the case of intermolecular NQEs, the quantum fluctuations serve to weaken intermolecular interactions (e.g., hydrogen bonds), thus reducing cohesion and the liquid density. Intramolecular NQEs, on the other hand, lead to longer average bond lengths with an associated increase in the molecular electrostatic moments and dispersion coefficients, which strengthens intermolecular interactions, effectively increasing liquid density.[58–60] The overall temperature dependence of the total NQEs results from the fact that intermolecular NQEs dominate at low temperatures, but decay quickly as the temperature increases, whereas intramolecular NQEs are not very sensitive to temperature.[58]

Comparison of DES-XFF predictions to experimentally measured hydrogen-deuterium isotope effects in molar volume, not just for water but also for other molecules (e.g., the difference between benzene isotopologues $C_6H_6$ and $C_6D_6$), provides an important additional validation of our models. Fig. 3G illustrates that for ethylene at 104 K and 169 K, DES-XFF predicted a stronger hydrogen-deuterium isotope effect than was measured experimentally, but that DES-XFF agreed with experiment to within the uncertainties for room-temperature methanol, acetone, and benzene. Results from first-principles force fields have rarely been quantitatively compared to experimentally measured hydrogen-deuterium isotope effects for molar volume in organic liquids.[61] Recent simulation work using a first-principles force field (TAFFI),[20] however, reported hydrogen-deuterium isotope effects for 92 organic liquids (finding an average decrease in molar volume of 0.73% and a maximum decrease of 2.3%), including acetone and methanol, for which comparisons can be made to both DES-XFF and experiment. For acetone, the molar-volume decrease of 0.93% predicted with TAFFI substantially exceeds the experimental value of 0.22% (compared with 0.31% predicted by DES-XFF); similarly, the corresponding TAFFI value for methanol (0.94%) is also larger than suggested by experiment or DES-XFF.

NQEs from simulations have been more commonly reported—as we have done in panels B and E of Fig. 3—as differences between properties evaluated using classical dynamics and using PIMD (with hydrogen atoms assigned a mass of one atomic mass unit), even though such differences cannot be measured experimentally. In one extensive study, the ARROW force field, a first-principles force field whose parameters have been developed for a wide range of substances,[18,19,62,63] was used to estimate the magnitude of NQEs for multiple liquids, and large effects were found for some properties, even at biologically relevant temperatures (for example, density changes of magnitude ~6%–8% for three alkanes in simulations at 272 K or 298 K[18]). More recently, an average increase in molar volume at room temperature of 1.9% across 92 organic liquids, with a maximum of 5.5%, was reported using TAFFI.[20] These effects are considerably smaller than the largest effects reported using ARROW, but are larger on average than the 0.6% density correction predicted by DES-XFF for the substances studied here at biologically relevant temperatures. Differences in the sets of molecules investigated in the different studies make it hard to draw definitive conclusions from comparing average behavior, but the data strongly suggest that DES-XFF predicts substantially more modest density changes from NQEs than do the other force fields considered here. That the magnitude of the density changes predicted by DES-XFF is reasonable is supported by the direct comparison with the experimentally measurable isotope effects discussed above.

## Solvation free energies

We next used DES-XFF to predict solvation free energies ($\Delta G_{solv}$), an important class of condensed-phase properties that contribute to the more complex protein-ligand binding free energies used to characterize affinities of drug molecules for their targets. We were interested in determining the accuracy of the $\Delta G_{solv}$ predictions as well as characterizing the contributions from both NQEs and MBD to these properties. Computing $\Delta G_{solv}$ requires not only an accurate and transferable model, but also one that is amenable to efficient implementation so that sufficient sampling can be achieved—a task that is further complicated when NQEs and MBD are treated explicitly. Using DES-XFF, we computed $\Delta G_{solv}$ values for 18 solute-solvent pairs: 12 values for aqueous solvation (water as solvent) and 6 values for solvation in benzene, using both PIMD and classical MD. The solutes ranged from very hydrophilic (e.g., NMA and imidazole, with experimental $\Delta G_{solv}$ values of about −10 kcal/mol) to hydrophobic (e.g., ethane and ethylene, with positive $\Delta G_{solv}$ values).

As shown in Fig. 4, DES-XFF achieved very good agreement with experiment, with MAEs of 0.29 kcal/mol for classical MD and 0.36 kcal/mol for PIMD. The largest discrepancy between DES-XFF predictions and experimental solvation free energies was observed for NMA in water, for which the PIMD and classical MD predictions of −8.9 and −9.1 kcal/mol, respectively, are less negative than the commonly cited experimental value of −10.0 kcal/mol.[64,65] It is worth noting, however, that a more recent compilation of experimental partition coefficients[66] yields an estimate of −9.4 kcal/mol for the hydration free energy of NMA at room temperature, in considerably better agreement with both DES-XFF predictions.

DES-XFF predicted only a small contribution of NQEs to solvation free energies, as evidenced by very similar results from PIMD and classical MD simulations. The average difference in $\Delta G_{solv}$ due to NQEs for the 12 solutes in water was only 0.17 kcal/mol (the average unsigned difference is also 0.17 kcal/mol), which appears consistent both with the near-zero experimental values for $\Delta G_{solv}$ of transferring benzene, ethane, acetone, and phenol from $H_2O$ to $D_2O$, and with the 0.1 kcal/mol experimental value for the isotope effect in water self-solvation $\Delta G_{solv}$.[67] This result contrasts with the 0.7 kcal/mol average difference in $\Delta G_{solv}$ due to NQEs obtained from the values reported for the same 12 solutes in water using the ARROW force field.[19] Another recent study,[68] using the empirical Q-AMOEBA water model with the adaptive quantum thermal bath method, reported values that correspond to an average NQE contribution to $\Delta G_{solv}$ of 0.50 kcal/mol for 38 solutes (0.37 kcal/mol for the eight solutes also studied with DES-XFF, for which DES-XFF predicted an average of 0.15 kcal/mol). These values are smaller than the ARROW result, but still substantially larger than the effect obtained in this work, which, as noted above, is consistent with expectations based on experimentally observed isotope effects. DES-XFF also predicted small differences in $\Delta G_{solv}$ due to NQEs for the 6 solutes in benzene, with an average magnitude of 0.04 kcal/mol and a signed average of −0.04 kcal/mol. On the other hand, we found that MBD contributions were more substantial, as they were for the density and $\Delta H_{vap}$, shifting $\Delta G_{solv}$ toward more positive values, by 0.30 kcal/mol on average in both solvents (Fig. 5; Table VII.II of the SI) and by as much as ~0.5 kcal/mol for aromatic solutes and NMA. It is worth noting, however, that the MBD contributions were not very sensitive to the solvent (Fig. 4B), and thus did not contribute significantly to the transfer free energy between water and benzene.

## Effect of QM data accuracy on condensed-phase predictions

As noted above, the development of DES-XFF relied on a large amount of coupled-cluster and MP2-level QM data. An important question is what would have happened if less accurate but faster QM methods had been used instead. In this section, we quantify the impact on calculated condensed-phase properties of replacing the most expensive QM calculations—non-covalent interaction energies of dimers, calculated using coupled-cluster theory with single, double, and perturbative triple excitations [CCSD(T)][69] or a local variant based on pair natural orbitals[70]—with more affordable but slightly less accurate calculations using an empirically fit density functional. Specifically, we recomputed the dimer interaction energies for neat water and acetone, and dimer and multimer nonadditive interaction energies for benzene, using ωB97M-V/def2-TZVPPD,[71] and refit our DES-XFF models to this new data. In most cases, the resulting DES-XFF variants showed a substantial degradation in the quality of the condensed-phase predictions (see Fig. VII.3 in the SI). The most straightforward variation, replacing the coupled-cluster dimer energies with DFT dimer energies, but retaining the MP2 multimer nonadditive energies and retaining MBD, resulted in liquids that were substantially less dense and less cohesive than in experiment for systems involving water, acetone, and benzene (Fig. VII.4 in the SI). This experiment highlights the importance of using accurate QM data in the development of first-principles models. It does not exclude the possibility that, through cancellation of errors, less accurate (and cheaper) data could be used successfully (see Fig. VII.4 in the SI for an example of an error cancellation that is successful for a limited set of systems), but we believe that the scope to reduce the accuracy of the QM data is limited. In contrast, there may be considerable scope to simplify DES-XFF's high-capacity functional form while still obtaining sufficiently accurate fits to the QM data to yield good condensed-phase results.

One deficiency of the DES-XFF water model is that it underestimates the enthalpy of vaporization by approximately 0.3 kcal/mol. Consistent with this underestimation, the oxygen-oxygen radial distribution function exhibits somewhat less structure than observed experimentally, particularly in the vicinity of the first minimum and second peak, and the self-diffusion coefficient is overestimated by approximately 20% (see Table VII.I and Fig. VII.1 in the SI). One possible explanation for at least a portion of these deviations is that our approach to estimating coupled-cluster-level water-water dimer interaction energies, in which MP2 energies extrapolated to the complete basis set limit are supplemented with a coupled-cluster correction computed using canonical CCSD(T) with the aug-cc-pVTZ[72] basis set, slightly underestimates the strength of intermolecular interactions. High-level quantum-chemical estimates[73] suggest that corrections beyond the frozen-core CCSD(T) approximation (which we have used in all our correlated QM calculations) increase the binding energy of the water dimer by approximately 1%. To evaluate the impact from this effect, we refitted the DES-XFF water model after scaling all reference water-water dimer interaction energies accordingly and found that the resulting model significantly improved agreement with experiment for density, enthalpy of vaporization, radial distribution functions, and self-diffusion coefficient.

## Conclusions

We show here that DES-XFF, a high-capacity first-principles force field, predicts condensed-phase properties of neat liquids and binary mixtures that are in most cases in excellent agreement with experiment across a range of small neutral molecules with different chemistries. Our results demonstrate that this level of accuracy depends not only on the high quality of the underlying QM data and the ability to fit that data with a high-capacity functional form, but also on accounting for NQEs through the use of PIMD, and especially on the inclusion of MBD, which was accomplished using an explicit force-field term. An important implication of our results is that, although additional physical effects could have been incorporated into the models, the current level of complexity appears sufficient to achieve a high level of accuracy, and is likely to be sufficient to obtain accurate condensed-phase results for other small neutral organic molecules.

Several limitations of this work should be noted. Agreement with experiment was less close for some properties than for liquid densities and vaporization enthalpies; the density of ice Ih was underestimated by 1.5%–1.7%, the self-diffusion coefficient of water was overestimated by approximately 20% (Table VII.I of the SI), and the excess volumes of methanol/benzene and pyrrole/benzene were overestimated by up to 0.4 and 1.1 $cm^3/mol$, respectively (Fig. VII.2 of the SI). Additionally, DES-XFF models have thus far been developed only for small, relatively rigid, neutral molecules composed of C, H, N, O, and S. Most importantly, the DES-XFF models used for the results presented here were parameterized separately for each neat liquid and each binary mixture. We have preliminary results with a simplified DES-XFF model—in which a molecule is described by the same set of parameters regardless of whether it is in a pure liquid or mixture—suggesting that system-specific parameterization may not be essential for achieving high accuracy. Nevertheless, extending DES-XFF to biomolecules would require both the development of transferable parameters and the generation of substantially larger QM datasets, which remain costly to obtain at the coupled-cluster level.

The ability of DES-XFF to simultaneously reproduce liquid densities and the much smaller experimentally observed differences in number density between isotopologues—on the order of a few tenths of a percent at room temperature—for both water and several organic liquids, without fitting to experimental data or introducing isotopologue-specific parameters, provides a particularly stringent demonstration of the accuracy with which DES-XFF captures NQEs. It additionally suggests that DES-XFF reproduces condensed-phase properties without relying on a cancellation of errors between NQEs and other physical effects such as MBD. Our results for water also provide support for the notion of competing quantum effects;[16,58–60,74] the destabilizing role of *inter*molecular NQEs, which weaken hydrogen bonds, is compensated to a large degree by *intra*molecular NQEs, which result in longer O–H bonds (DES-XFF predicts that NQEs increase the average bond length by 0.016 Å in liquid $H_2O$ and 0.012 Å in $D_2O$) that strengthen intermolecular interactions. Capturing the isotope effect on the bond lengths is straightforward and requires only a well-parameterized anharmonic stretch potential. Accurately modeling the dependence of the intermolecular interactions on the bond lengths, however, has been a particular challenge in developing water force fields. DES-XFF's ability to accurately capture water properties suggests that it adequately models the coupling of these competing quantum effects.

We found that MBD plays a larger role than NQEs for most of the condensed-phase properties studied (Fig. 5). Our results suggest that MBD contributions are substantial enough that they should be accounted for when developing first-principles force fields, including machine-learned models. We note that the MBD term in DES-XFF is fit independently of the other terms, and a related approach could potentially be adopted by machine-learned models, which are typically fit only to data for monomers and dimers and sometimes small clusters, in which MBD effects are not prominent.

## Acknowledgements

The authors thank Zeph Arvanitis, Cory Hargus, and Kuang Yu for help with preliminary investigations; Panagiotis Angelikopoulos, Brent A. Gregersen, Je-Luen Li, Robert T. McGibbon, and Kim Palmo for general discussion of QM and force field issues; Asaph Azaria for help with resource planning; Asaph Azaria and Sarah Christensen for help with project coordination; and Eric Martens for editorial assistance.

## Figures

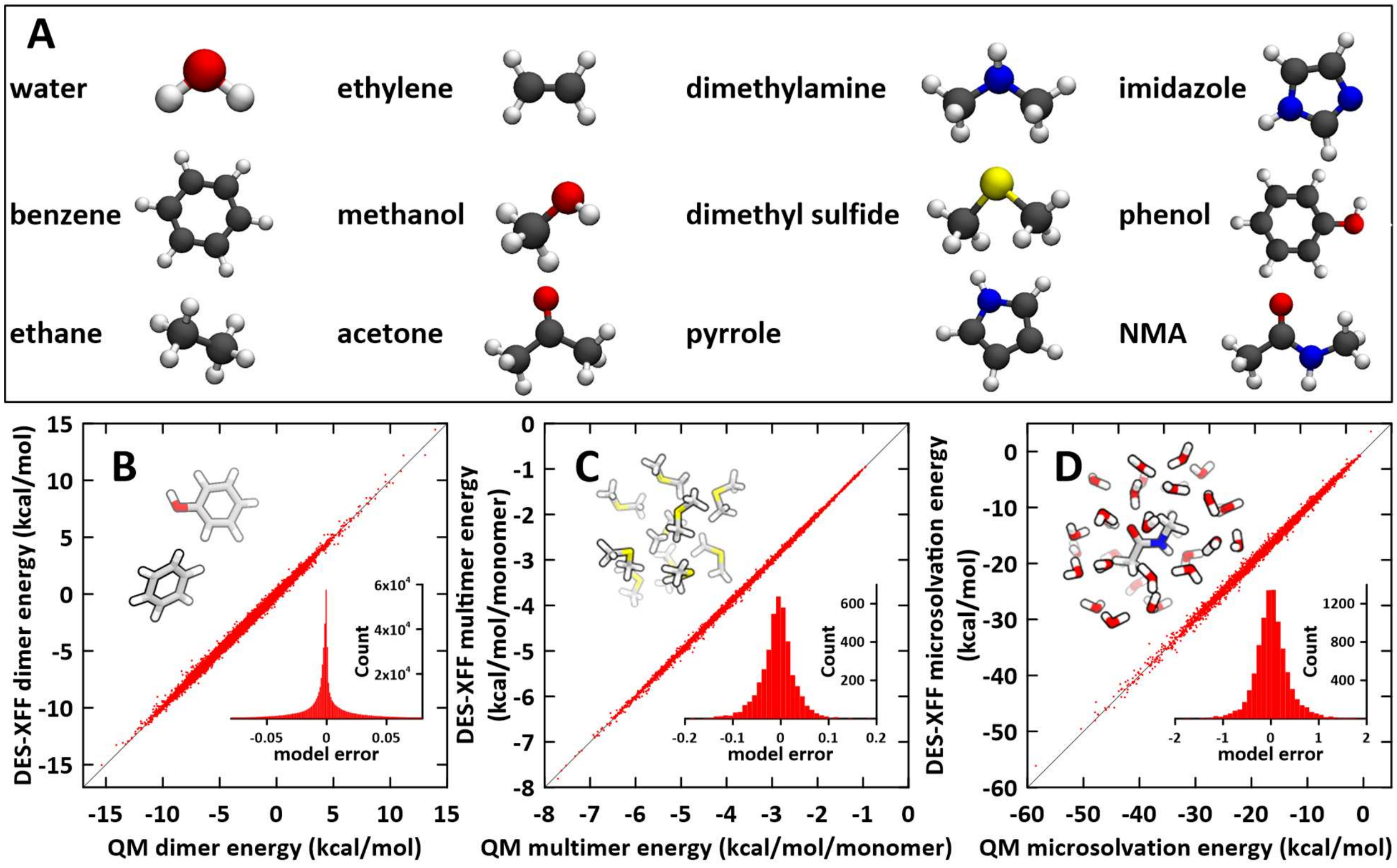

**Fig. 1. Comparison of DES-XFF versus QM interaction energies for all studied systems.** **(A)** Molecules studied in this paper. **(B)** Dimer interaction energies for 604,273 training and validation geometries (see section V.A of the SI). The top inset shows a representative dimer configuration, and the bottom inset shows the distribution of DES-XFF errors in dimer interaction energies. **(C)** Same as (B) but for multimer interaction energies divided by the number of molecules *n* in the multimer (4095 geometries; *n* varies from 7 to 40). **(D)** Same as (B) but for microsolvation energies, defined as the interaction energy of a molecule at the center of a large molecular cluster with the rest of the cluster (9054 geometries). The molecular clusters in panels (C) and (D) contain an average of 109 atoms (ranging from 60 to 272) and have an average radius of gyration of 5.5 Å (ranging from 3.9 to 9.0 Å). DES-XFF reproduces the reference QM data with mean absolute errors of 0.024 kcal/mol (3.1% of the mean absolute

QM energy) for dimer energies, 0.027 kcal/mol/monomer (0.8%) for multimer energies, and 0.26 kcal/mol (1.9%) for microsolvation energies.  All geometries were extracted from PIMD simulations with a preliminary force field to ensure adequate sampling of substantial intramolecular deformations arising from NQEs, which would be less frequently sampled in classical MD simulations.

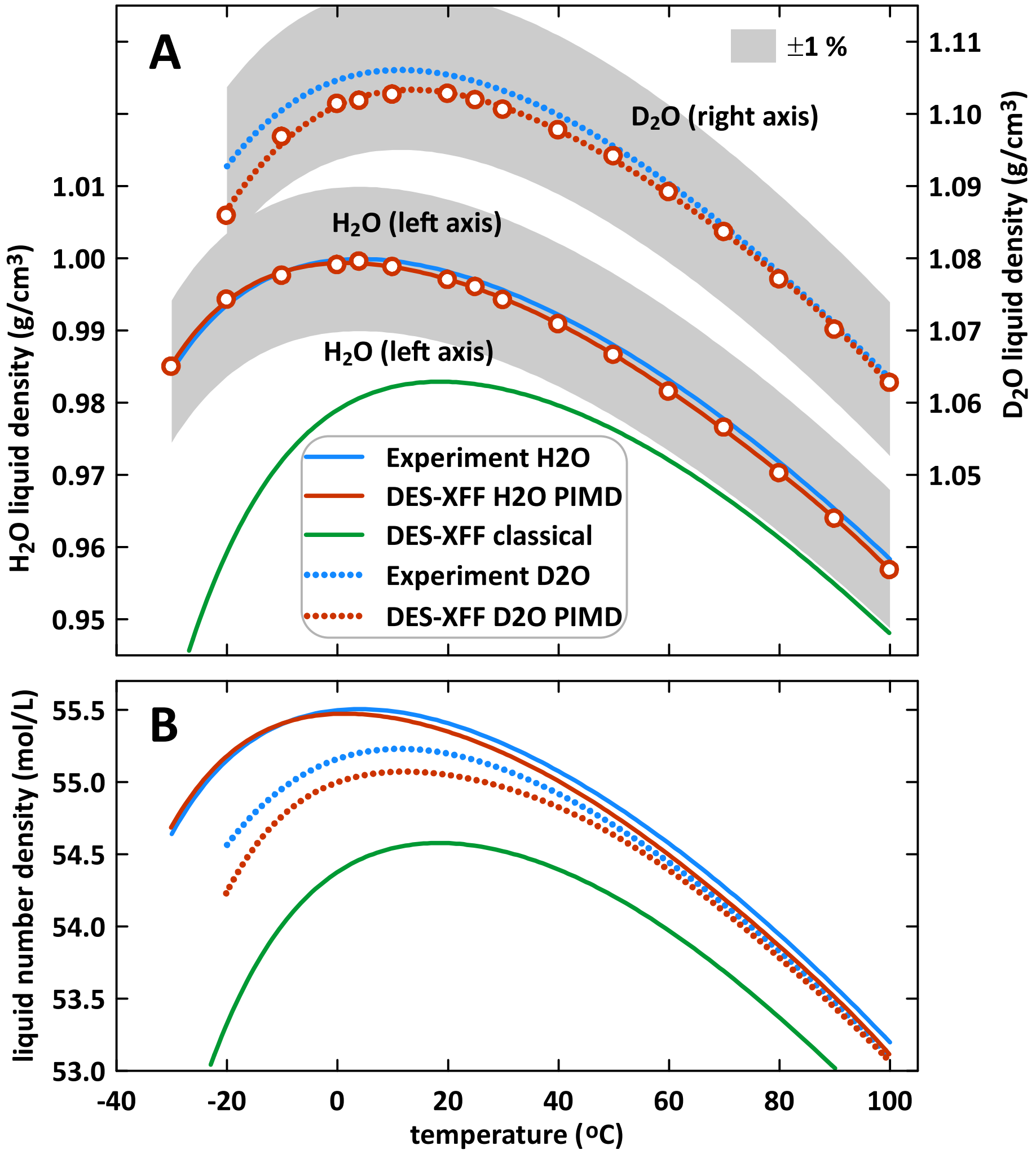

**Fig. 2. Predicted density of liquid water and its dependence on NQEs. (A)** Density as a function of temperature for liquid $H_2O$ and $D_2O$ computed with PIMD (MAE relative to experiment of 0.1% for $H_2O$ and 0.2% for $D_2O$) and with classical MD; experimental data is from references 75 ($H_2O$) and 76 ($D_2O$). As a visual aid, the regions within ±1% density deviation from experiment are shaded. **(B)** Same as (A) but for number density of liquid water.

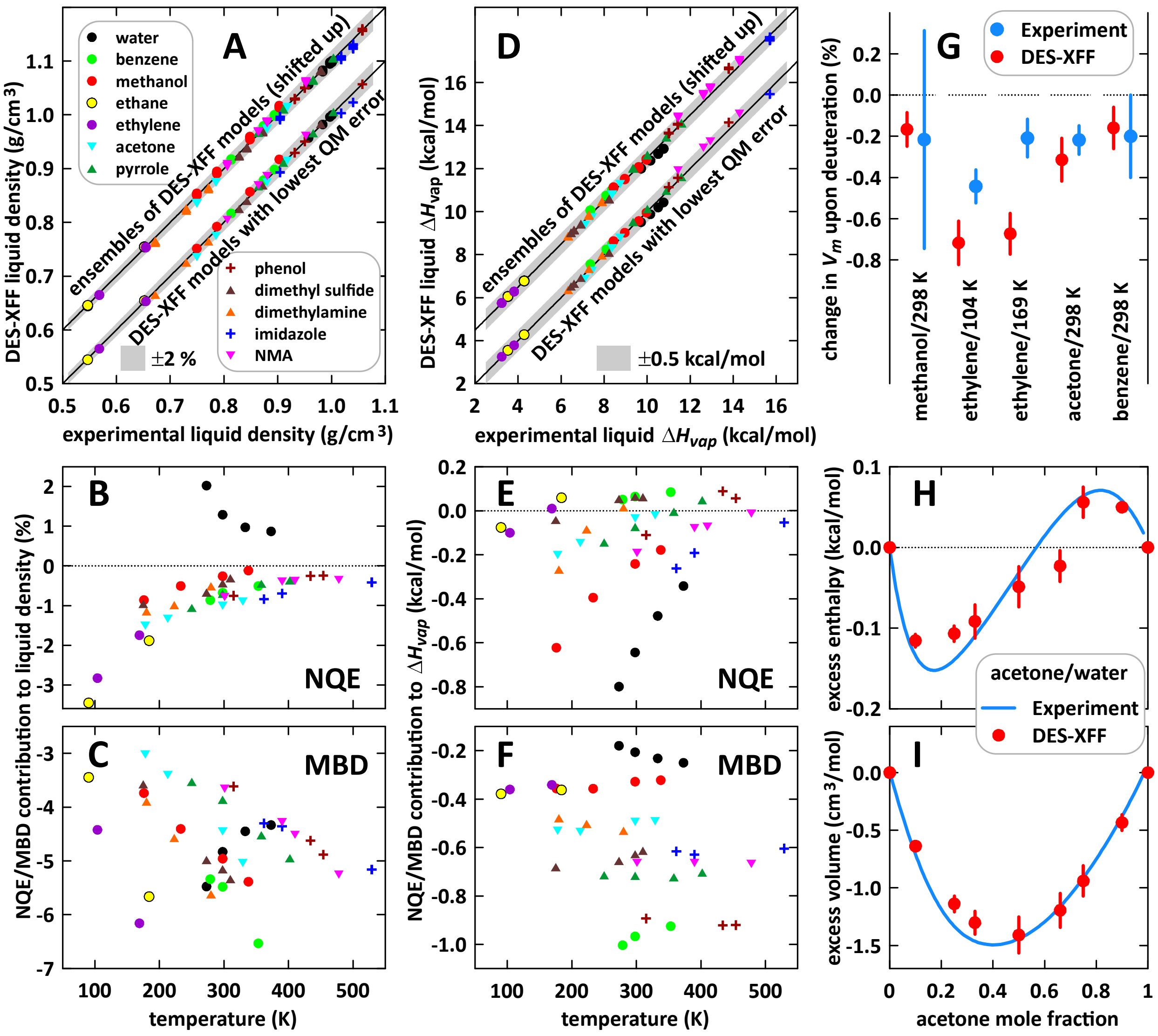

**Fig. 3. Predicted properties of organic liquids and mixtures from PIMD simulations. (A)** Comparison of DES-XFF predictions with experimental liquid densities at various temperatures between the melting and boiling points of each substance. The results are shown (i) for models with the lowest QM error (one per substance) and (ii) for ensembles of DES-XFF models (five per substance, though it is not possible to visually distinguish all the points, as the different members of a given ensemble typically gave very similar results). **(B)** DES-XFF predictions for differences in liquid density due to NQEs (PIMD minus classical). **(C)** Same as (B) but for MBD (PIMD with MBD minus PIMD without MBD). **(D)** Same as (A) but for vaporization enthalpy $\Delta H_{vap}$. **(E)** DES-XFF predictions for differences in $\Delta H_{vap}$ due to NQEs (PIMD minus classical). **(F)** Same as (E) but for MBD (PIMD with MBD minus PIMD without MBD). **(G)** Comparison of DES-XFF predictions with experimental values for the change of liquid molar volume upon complete deuteration (e.g., $C_6H_6$ to $C_6D_6$ for benzene). **(H)** Comparison of DES-XFF predictions with experimental values for the excess enthalpy of the acetone/water mixture at 298 K. **(I)** Same as (H) but for the excess volume. In panels (G), (H), and (I), DES-XFF values are averaged over the five corresponding models; the error bars correspond to the 95% confidence interval. As a visual aid, the shading in panels (A) and (D) denotes regions within ±2% density deviation and ±0.5 kcal/mol $\Delta H_{vap}$ deviation, respectively. In panels (B) through (F), symbols are as in panel (A).

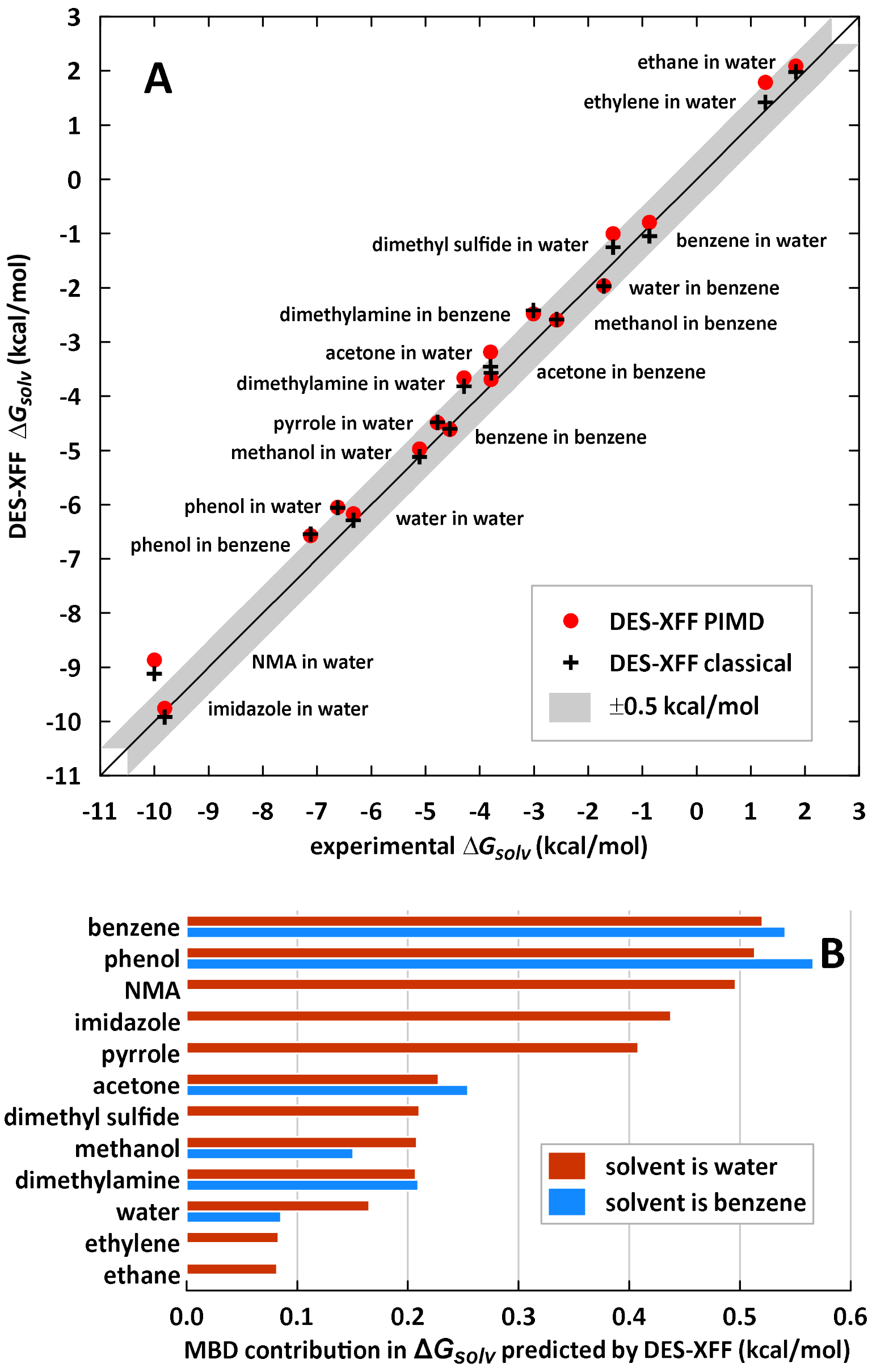

**Fig. 4. Predicted solvation free energies ($\Delta G_{solv}$). (A)** Comparison of DES-XFF predictions with experimental solvation free energies ($\Delta G_{solv}$).[65,77] The typical statistical uncertainties are

±0.1 and ±0.05 kcal/mol for the PIMD and classical simulations, respectively. As a visual aid, the shaded region indicates deviations of up to ±0.5 kcal/mol in $\Delta G_{solv}$. **(B)** DES-XFF predictions for the MBD contribution to $\Delta G_{solv}$.

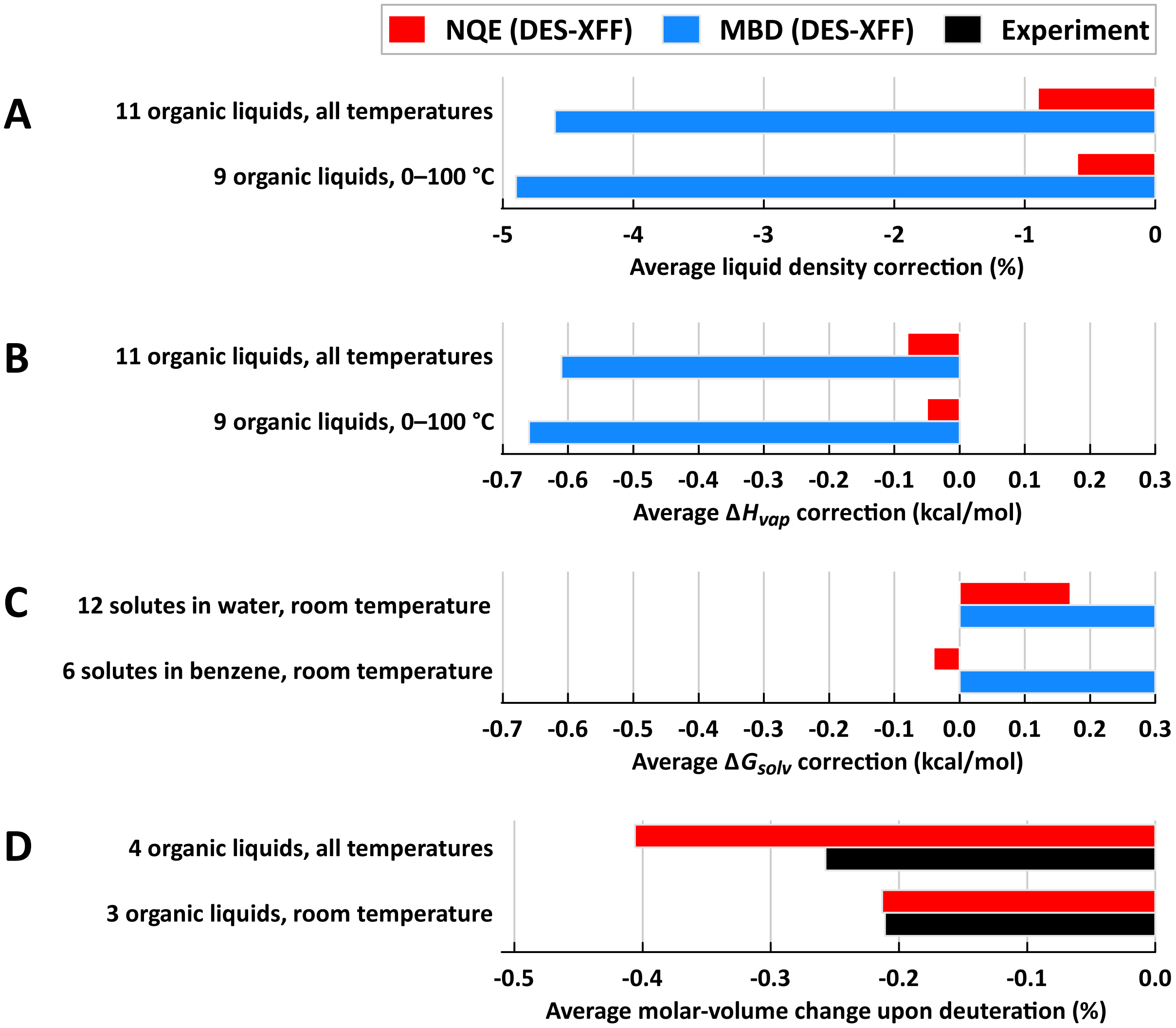

**Fig. 5. Average contributions of NQEs and MBD to condensed-phase properties predicted by DES-XFF. (A)** Average changes in liquid density due to NQEs and MBD for 11 organic liquids over all studied temperatures and for 9 of these liquids over the 273–373 K range (excluding ethane and ethylene, which boil below 273 K). **(B)** Same as (A), but for vaporization enthalpy ($\Delta H_{vap}$). **(C)** Average changes in solvation free energy ($\Delta G_{solv}$) due to NQEs and MBD for 12 solutes in water and 6 solutes in benzene at room temperature. **(D)** Comparison of DES-XFF predictions with experimental values for the average percentage change in molar volume upon complete deuteration of four organic liquids (acetone, benzene, methanol, and ethylene) over all studied temperatures and of three of these liquids (excluding ethylene) at room temperature. The all-temperature average in panel (D) includes five data points, with ethylene contributing at two temperatures (104 and 169 K). All values are signed averages over the indicated data points. In panels (A)–(C), NQE contributions are calculated as differences between PIMD and classical MD simulations, whereas MBD contributions are calculated as differences between PIMD simulations with and without MBD.

# Supplementary Information

# Highly accurate first-principles models of molecular interactions

This document is organized as follows:

- Sections I–IV describe the DES-XFF functional form, with I–III containing preliminary definitions and IV containing the energy terms
- Section V describes the fitting procedure
- Section VI provides details of the MD simulations, including PIMD
- Section VII contains additional condensed-phase results not included in the main text
- Section VIII describes the sources of the experimental data used to validate the DES-XFF models
- Sections IX and X contain details of the QM calculations

# I. Local reference axis systems (LORAS)

We used four types of LORAS, as described below. Each type of LORAS is associated with one or more symmetries, which are enforced for the relevant coupling coefficients (model parameters) for the local multipoles at the host atom. For example, symmetry with regard to inverting the *z*-axis implies that all coupling coefficients that are odd with regard to the *z* coordinate are required to be zero. Note that our LORAS definitions do not necessarily involve all nearest neighbors of the host atom. Below, we use $\boldsymbol{e}_{BA}$ as a shorthand notation for a unit vector $\boldsymbol{e}_{BA} = (\boldsymbol{r}_A - \boldsymbol{r}_B)/\|\boldsymbol{r}_A - \boldsymbol{r}_B\|$. As is conventional, $\boldsymbol{a} \times \boldsymbol{b}$ denotes the cross-product of two vectors. All axis vectors defined below are normalized to unit length after construction.

## *A. Two neighbors*

This type of LORAS (which is similar, but not identical, to System 3 in Ref. 1) we designate as `'water_o'`. It requires that the host atom $A$ has (at least) two chemically bonded neighbors $B_1$ and $B_2$.

$$\boldsymbol{z} = \boldsymbol{e}_{B_1A} \times \boldsymbol{e}_{B_2A}$$
$$\boldsymbol{y} = \boldsymbol{z} \times \left( \left( \boldsymbol{e}_{B_1A} + \boldsymbol{e}_{B_2A} \right) \times \boldsymbol{z} \right)$$
$$\boldsymbol{x} = \boldsymbol{y} \times \mathbf{z}$$

Symmetries:

- with regard to inverting the *z*-axis (always)
- with regard to inverting the *x*-axis, if $B_1$ and $B_2$ have the same atom type (e.g., in water)

### B. *One neighbor and one second neighbor*

This type of LORAS (which is similar, but not identical, to System 7 in Ref. 1) we designate as `'water_h'`. It requires that the host atom $A$ has (at least) one chemically bonded neighbor $B_1$ and (at least) one atom $B_2$ bonded to $B_1$, but not bonded to $A$.

$$\boldsymbol{z} = \boldsymbol{e}_{B_1 A}$$
$$\boldsymbol{y} = \boldsymbol{z} \times \boldsymbol{e}_{B_2 B_1}$$
$$\boldsymbol{x} = \boldsymbol{y} \times \mathbf{z}$$

Symmetries:

- with regard to inverting the *y*-axis (always)

### C. *One neighbor and two second neighbors*

This type of LORAS (which applies in the same environments as System 6 in Ref. 1, though it defines axes differently) we designate as `'benzene_h'`. It requires that the host atom $A$ has (at least) one chemically bonded neighbor $B_1$ and (at least) two atoms $B_2$ and $B_3$ bonded to $B_1$, but not bonded to $A$.

$$\boldsymbol{z} = \boldsymbol{e}_{B_1 A}$$
$$\boldsymbol{y} = \boldsymbol{z} \times \left(\boldsymbol{e}_{B_2 B_1} - \boldsymbol{e}_{B_3 B_1}\right)$$
$$\boldsymbol{x} = \boldsymbol{y} \times \mathbf{z}$$

Symmetries:

- with regard to inverting the *y*-axis (always)
- with regard to inverting the *x*-axis, if $B_2$ and $B_3$ have the same atom type (e.g., in benzene)

### *D. One neighbor*

This type of LORAS, which we designate as `'diatomic'`, requires that the host atom $A$ has (at least) one chemically bonded neighbor $B_1$.

$$\boldsymbol{z} = \boldsymbol{e}_{B_1A}$$
$$\boldsymbol{y} = \text{arbitrary} \perp \boldsymbol{z}$$
$$\boldsymbol{x} = \boldsymbol{y} \times \mathbf{z}$$

Symmetries:

- with regard to inverting the *x*-axis (always)
- with regard to inverting the *y*-axis (always)
- with regard to swapping the *x*- and *y*-axes (always)

These three symmetries correspond, for multipoles up to rank 3, to cylindrical symmetry with regard to the *z*-axis. Symmetry with regard to swapping the *x*- and *y*-axes implies that the coupling coefficients for moments $x^p y^q z^r$ and $x^q y^p z^r$ are required to be equal.

## II. Dynamic multipoles

Our models feature conformationally dependent (dynamic) multipoles,

*Equation 1*

$$q_{i\alpha}^X = \tilde{q}_{I\alpha}^X + \sum_{t \in \Theta_I} C_{I\alpha t}^X \Phi_{it},$$

where $X$ designates one of the three energy components—electrostatics ($X = \text{ES}$), exchange ($X = \text{EX}$), or dispersion ($X = \text{DS}$); $i$ is the atomic index; $I$ is the type of atom $i$; $\alpha$ is a multipole component (monopole, x, y, z, xx, xy, xz, yy, …, zzz); $\tilde{q}_{I\alpha}^X$ is the static value of the multipole (model parameter); $\Theta_I$ designates the set of geometry factors (GFs) associated with the atom type $I$; $\Phi_{it}$ is the $t$-th GF for atom $i$; and $C^{X}_{I\alpha t}$ is the coupling coefficient (model parameter). Importantly, Equation 1 defines *local* multipoles (i.e., referenced to the LORAS of atom $i$). To compute the interaction energy, all local multipoles are transformed into the global frame.

We consider two types of GFs: intramolecular and intermolecular.

### *A. Intramolecular geometry factors*

Intramolecular GFs have the following form:

*Equation 2*

$$\Phi_{it} = \sigma_t \tanh\left[\left(\sum_{j\in\Xi_{it}} x_{j,i}^{p_t^x} y_{j,i}^{p_t^y} z_{j,i}^{p_t^z} - \mu_t\right)/\sigma_t\right]$$

where $\Xi_{it}$ is the set of atoms (either immediate neighbors of atom $i$ or its second neighbors) specified in the definition of the GF; $x_{j,i}, y_{j,i}, z_{j,i}$ are the coordinates of atom $j$ in the LORAS of atom $i$; $p_t^x, p_t^y, p_t^z$ are integer powers (0 or 1) specified in the definition of the GF; and offset $\mu_t$ and dynamic range $\sigma_t$ are precomputed hyperparameters. The offset $\mu_t$ and dynamic range $\sigma_t$ were set to the mean and three times the standard deviation, respectively, of the coordinate-dependent sum in Equation 2, evaluated over the training geometries.

### *B. Intermolecular geometry factors*

Intermolecular GFs have the following form:

*Equation 3*

$$\Phi_{it} = \sum_{j\in\Psi_{it}} \tanh\left(e^{\beta_t(R_t - r_{ij})}\right) \zeta_{m_t}(i;\mu_t,\sigma_t)\, \zeta_{n_t}(j;\mu_t,\sigma_t) s\left(r_{ij}/R_{max}\right)$$

$$\zeta_0(a;\mu,\sigma) = 1, \qquad \zeta_1(a;\mu,\sigma) = \sum_{b\ bonded\ to\ a} \sigma\, tanh[(r_{ab} - \mu)/\sigma],$$

where $\Psi_{it}$ is the set of atoms (belonging to molecules other than the one containing atom $i$) specified in the definition of the GF; $r_{ij}$ is the interatomic separation; $\beta_t$, $R_t$, and $R_{\max}$ are hyperparameters (we use $\beta_t$ = 1 bohr$^{-1}$ and $R_{\max}$ = 7 bohr for all GFs); $m_t$ and $n_t$ are integer indices (0 or 1) specified in the definition of the GF; and offset $\mu_t$ and dynamic range $\sigma_t$ are additional hyperparameters. The switching function is

Equation 4

$$s(x) = \begin{cases} 1 & if \quad x < 0 \\ cos^2 \frac{\pi}{2} x & if \;\; 0 \leq x < 1 \\ 0 & if \; x \geq 1 \end{cases},$$

as has been used by others, for example in Ref. 2.

## III. Interaction kernels

In our model, each atom is characterized by several sets of nonredundant Cartesian multipole components up to octupoles (maximum rank 3; 20 components in total),

$$q(0,0,0), q(1,0,0), q(0,1,0), q(0,0,1), q(2,0,0), q(1,1,0), \dots, q(0,0,3)\,,$$

where, for example, $q(0,0,0)$ is the monopole, $q(1,0,0)$ is the x-component of the dipole, $q(1,1,0)$ is the xy-component of the quadrupole, and $q(0,1,2)$ is the yzz-component of the octupole. Note that not every set of multipoles used in DES-XFF includes all 20 components. In particular, point electrostatic charges, both static and dynamic, are associated with only one component—the monopole; diffuse static multipoles include 10 components—monopole, dipole, quadrupole; induced dipoles include 3 components—$q(1,0,0), q(0,1,0), q(0,0,1)$.

The interaction energy between two sets of Cartesian multipoles is

Equation 5

$$U_{ij}^X = \sum_{\alpha,\beta} \frac{q_i^X\left(k_\alpha^x, k_\alpha^y, k_\alpha^z\right) q_j^X\left(k_\beta^x, k_\beta^y, k_\beta^z\right)}{k_\alpha^x!\, k_\alpha^y!\, k_\alpha^z!\, k_\beta^x!\, k_\beta^y!\, k_\beta^z!} \frac{d^{k_\alpha^x+k_\alpha^y+k_\alpha^z+k_\beta^x+k_\beta^y+k_\beta^z}}{dx_i^{k_\alpha^x} dy_i^{k_\alpha^y} dz_i^{k_\alpha^z} dx_j^{k_\beta^x} dy_j^{k_\beta^y} dz_j^{k_\beta^z}} F^X\left(\|\boldsymbol{r}_i - \boldsymbol{r}_j\|; w_i, w_j\right),$$

where $i$ and $j$ are two distinct atomic sites with coordinates $\boldsymbol{r}_i = (x_i, y_i, z_i)$ and $\boldsymbol{r}_j = (x_j, y_j, z_j)$; $w_i, w_j$ are width parameters; and $F^X$ is an interaction kernel function for one of the three energy components: electrostatics ($X = \text{ES}$), exchange ($X = \text{EX}$), or dispersion ($X = \text{DS}$). Importantly, multipoles $q_i$ and $q_j$ in Equation 5 are referenced to the same global frame.

### *A. Electrostatic kernel*

The electrostatic kernel for two diffuse exponential charges is given by (as shown in Ref. 3)

Equation 6

$$F^{ES}\left(r; w_i, w_j\right) = \frac{1}{r}\left[1 - f\left(r; w_i, w_j\right) e^{-r/w_i} - f\left(r; w_j, w_i\right) e^{-r/w_j}\right],$$

$$\text{where } f(r; u, v) = \frac{u^4(3v^2 - u^2)}{(v^2 - u^2)^3} + r\frac{u^3}{2(v^2 - u^2)^2}$$

If one or both charges have zero width (point charges), the kernel simplifies to

$$F^{ES}(r; w, 0) = \frac{1}{r}\left[1 - \left(1 + \frac{r}{2w}\right) e^{-r/w}\right],$$

$$\text{and } F^{ES}(r; 0,0) = \frac{1}{r},$$

the latter being the usual Coulomb kernel for the multipole expansion of point electrostatics.

If the two charges have equal widths, the kernel reduces to

$$F^{ES}(r;w,w)=\frac{1}{r}\left[1-\left(1+\frac{11}{16}\left(\frac{r}{w}\right)+\frac{3}{16}\left(\frac{r}{w}\right)^2+\frac{1}{48}\left(\frac{r}{w}\right)^3\right)e^{-r/w}\right].$$

### *B. Exchange kernel*

The exchange kernel (identical to the one used in Ref. 3) is

*Equation 7*

$$F^{EX}(r;w_i,w_j)=\frac{e^{-r/w_j}-e^{-r/w_i}}{r/w_i-r/w_j}.$$

In the case of two equal widths, the kernel simplifies to

$$F^{EX}(r;w,w)=e^{-r/w}.$$

### *C. Dispersion kernel*

The dispersion kernel is

$$F^{DS}(r;w_i,w_j)=-\frac{1}{(r/w_{ij})^6+a_6^6}-\frac{1}{(r/w_{ij})^8+a_8^8},$$

where $w_{ij}=(w_i+w_j)/2$,

and $a_6$ and $a_8$ are hyperparameters; we use $a_6=1$ and $a_8=1$.

## IV. Energy

We consider three categories of multipoles (see sections IV.A, IV.B, and IV.C below for specification of which components from ES, EX, and DS are relevant to each category):

i. dynamic multipoles (labeled “d”) up to octupoles, interacting only with other dynamic multipoles
ii. static multipoles (labeled “s”) up to quadrupoles, interacting only with the induced dipoles
iii. induced dipoles (labeled “i”), interacting with both the static multipoles and other induced dipoles

Note that for the electrostatic interactions, our model also includes point charges at each atom (representing the nuclei), for both the dynamic and static model components. Values of the point charges $q_i^{\mathrm{n}}(0,0,0)$ are not independent model parameters; they are obtained for the dynamic and static model components as

Equation 8

$$q_i^n(0,0,0) = \sum_{b \in bonds} \mathcal{M}_{ib} q_b - q_i^c(0,0,0)\,,$$

where $q_i^{\mathrm{c}}(0,0,0)$ is the value of the diffuse monopole for that model component (aside from the nuclear charges, all electrostatic multipoles in DES-XFF are diffuse; “c” stands for “cloud”) and $q_b$ is a bond-charge increment (BCI) along the $b$-th bond. $\mathcal{M}$ is a matrix whose elements take the values −1, 0, or 1 and which encodes the topology of the molecule; rows correspond to the atoms and columns correspond to the bonds. Each column in $\mathcal{M}$ contains exactly two nonzero entries (1 and −1) corresponding to the atoms connected by the bond. Note that in the case of dynamic charges, we fit coupling coefficients (see Equation 1) for the atomic BCI-propensities $\mathscr{p}_i$, from which the BCIs are computed as

Equation 9

$$q_b = \sum_{i \in atoms} \mathcal{M}_{bi}^T \mathscr{p}_i \,,$$

where $\mathcal{M}^{\mathrm{T}}$ is the transpose of matrix $\mathcal{M}$. In the case of static charges, BCIs are the model parameters, whereas in the case of dynamic charges, the BCIs are not independent and are computed from the atomic BCI-propensities.

The total energy of a system can be written as a sum of the dynamic-dynamic $U^{\mathrm{dd}}$, static-induced $U^{\mathrm{si}}$, induced-induced $U^{\mathrm{ii}}$, core $U^{\mathrm{c}}$, induction restraint $U^{\mathrm{r}}$, valence $U^{\mathrm{v}}$, and many-body dispersion $U^{\mathrm{mbd}}$ terms

Equation 10

$$U = \sum_{i \in A, j \in B, A<B} \left(U_{ij}^{dd} + U_{ij}^{si} + U_{ij}^{ii} + U_{ij}^{c}\right) + \sum_{i} U_i^r + \sum_{A} U_A^v + \sum_{i \in A, j \in B, k \in C, A<B<C} U_{ijk}^{mbd} \,,$$

where $i, j, k$ are atomic indices, $A, B, C$ are molecular indices, $\{i \in A,\ j \in B, A < B\}$ designates all distinct *intermolecular* pairs of atoms, and $\{i \in A,\ j \in B, k \in C, A < B < C\}$ designates all distinct *intermolecular* triplets of atoms such that no pair of them belongs to the same molecule.

## *A. Dynamic-dynamic interaction energy*

The dynamic-dynamic interaction energy for an intermolecular pair of atoms is

*Equation 11*

$$U_{ij}^{dd} = U_{ij}^{dd,ES,nn} + U_{ij}^{dd,ES,nc} + U_{ij}^{dd,ES,cn} + U_{ij}^{dd,ES,cc} + U_{ij}^{dd,EX} + U_{ij}^{dd,DS} .$$

Note that Equation 11 includes four electrostatic terms: nucleus-nucleus (nn), nucleus-cloud (nc, cn), and cloud-cloud (cc), where "nucleus" corresponds to the point atomic charge, and "cloud" corresponds to all diffuse multipoles up to octupole.

## B. Static-induced energy

The static-induced energy for an intermolecular pair of atoms is electrostatic only and is

*Equation 12*

$$U_{ij}^{si} = U_{ij}^{si,ES,nd} + U_{ij}^{si,ES,cd} + U_{ij}^{si,ES,dn} + U_{ij}^{si,ES,dc} .$$

The terms in Equation 12 are nucleus-dipole (nd, dn) and cloud-dipole (cd, dc).

## C. Induced-induced energy

The induced-induced energy for an intermolecular pair of atoms is the interaction between the two induced dipoles

*Equation 13*

$$U_{ij}^{ii} = U_{ij}^{ii,ES,dd} + \varepsilon U_{ij}^{ii,EX,dd} ,$$

where $\varepsilon$ is 0 or 1.

### *D. Core energy*

The core energy for an intermolecular pair of atoms is

*Equation 14*

$$U_{ij}^{c} = \epsilon e^{-\rho_{ij}} s\left(\rho_{ij}/k\right), \quad \rho_{ij} = (r_{ij} - d_{IJ})/w \,,$$

where $r_{ij}$ is the interatomic separation and $d_{IJ} = (d_I + d_J)/2$ is a characteristic distance that depends on the types $I$ and $J$ of atoms $i$ and $j$, respectively. $\epsilon = 0.001$ au, $k = 2$, and $w = 0.2$ au are universal constants for all atom types. The $d_I$ parameters are not subject to fitting; their values (2 au for hydrogens and 3 au for heavy atoms) are such that the core energy is zero or negligible for all geometries in the training/validation set.

The switching function (which has been used previously, for example in Ref. 2) is

*Equation 15*

$$s(x) = \begin{cases} 1 & if \quad x < 0 \\ cos^2 \frac{\pi}{2} x & if \;\; 0 \le x < 1 \\ 0 & if \; x \ge 1 \end{cases} \,.$$

### *E. Induction restraint energy*

The induction restraint energy for atom $i$ is

*Equation 16*

$$U_i^r = \mu^2 \left(1 - \sqrt{1 - \boldsymbol{d}_i^T \boldsymbol{\alpha}_i^{-1} \boldsymbol{d}_i / \mu^2}\right),$$

where $\boldsymbol{d}_i$ is the induced dipole (3-vector determined by the conformation of the system) and $\boldsymbol{\alpha}_i$ is the atomic polarizability (3 × 3 matrix of model parameters); $\mu = 0.3$ hartree$^{1/2}$ is an atom-independent hyperparameter. In the limit of small induced dipoles, Equation 16 reduces to the conventional polarization self-energy ½ $\boldsymbol{d}^T\boldsymbol{\alpha}^{-1}\boldsymbol{d}$. As $\boldsymbol{d}^T\boldsymbol{\alpha}^{-1}\boldsymbol{d}$ approaches $\mu^2$, the slope of the restraint potential diverges, thereby imposing an upper bound on the magnitude of the induced dipole for a given atomic polarizability. This is a modified version of the restraint potential from Ref. 3.

The induced dipoles are determined by minimizing the sum of the static-induced, induced-induced, and induction restraint energy terms $U^{si} + U^{ii} + U^r$ with respect to the induced dipole moments. Polarization in DES-XFF is purely intermolecular; neither the static-induced nor the induced-induced energy includes intramolecular interactions. In MD simulations, the induced-dipole degrees of freedom are handled using the extended Lagrangian approach described in section VI.

### *F. Valence energy*

The valence energy for molecule $A$ is:

Equation 17

$$U_A^v = \sum_{t\in\Upsilon_A} C_t \prod_{\{\alpha,n\}\in\Omega_t} \phi_\alpha(n)\,,$$

where $\Upsilon_A$ is the set of all valence terms (*all* stretch and bend diagonal terms up to $n = 4$, *select* torsion diagonal terms up to $n = 3$, and *select* cross-terms with cumulative orders 2 and 3), $\Omega_t$ is a set of internal coordinates $\phi_\alpha$ with the corresponding orders $n$ involved in the $t$-th term, and $C_t$ is a linear coefficient (model parameter). Note that in the special case when $\Omega_t$ is an empty set (the free term), the product is interpreted as equal to 1.

We consider three types of internal coordinates:

i. **stretch**: $\phi_\alpha(n) = (r_\alpha - r_\alpha^0)^n$, where $r_\alpha$ is the corresponding bond length and $r_\alpha^0$ is an offset constant

ii. **bend**: $\phi_\alpha(n) = (\cos\theta_\alpha - \cos\theta_\alpha^0)^n$, where $\theta_\alpha$ is the corresponding angle and $\theta_\alpha^0$ is an offset constant

iii. **torsion**: $\phi_\alpha(n) = \cos n\chi_\alpha - \cos n\chi_\alpha^0$, where $\chi_\alpha$ is the corresponding dihedral angle and $\chi_\alpha^0$ is an offset constant

### *G. Many-body dispersion energy*

The many-body dispersion energy for a triplet of atoms belonging to three distinct molecules is given by a damped Axilrod–Teller–Muto (DATM) potential

*Equation 18*

$$U_{ijk}^{mbd} = C_{IJK} \frac{1 + 3\cos\theta_i \cos\theta_j \cos\theta_k}{\left(r_{ij}^3 + w_{IJ}^3\right)\left(r_{jk}^3 + w_{JK}^3\right)\left(r_{ki}^3 + w_{KI}^3\right)} H\left(R_{cut} - max\left(r_{ij}, r_{jk}, r_{ki}\right)\right),$$

where $r_{ij}$ are interatomic separations; $\theta_i$ are angles in the triangle formed by atoms $i, j, k$; $I, J, K$ are the types of atoms $i, j, k$, respectively; and $w_{IJ} = (w_I + w_J)/2$, where the damping widths $w_I$ are model parameters. (The original ATM potential can be found in Ref. 4.) The coefficient $C_{IJK}$ is given by

$$C_{IJK} = \frac{3}{2}\alpha_I\alpha_J\alpha_K \frac{\omega_I\omega_J\omega_K\left(\omega_I + \omega_J + \omega_K\right)}{\left(\omega_I + \omega_J\right)\left(\omega_J + \omega_K\right)\left(\omega_K + \omega_I\right)},$$

where $\alpha_I$ and $\omega_I$ are model parameters. The parameters $\alpha_I$ and $\omega_I$ are independently optimized DATM model parameters and are distinct from the corresponding parameters of the CFDM model used to generate the reference MBD energies (see section V.C). For all hydrogen atoms, we set $\alpha_I = 0$; as a result, only triplets of heavy atoms contribute to the many-body dispersion energy.

$H(x)$ is the Heaviside step function

$$H(x) = \begin{cases} 1, & x > 0 \\ 0, & x \leq 0 \end{cases},$$

which ensures that only relatively compact triplets of atoms contribute to the many-body dispersion energy; we use $R_{\text{cut}} = 15.1178$ au $= 8$ Å. The DATM parameters were optimized with this cutoff in place, so the effect of truncating the longer-range contributions is implicitly accounted for in the fitted parameters.

## V. Fitting procedure

This section describes our protocols for obtaining model parameters. A system is defined as a neat liquid or a binary system consisting of two distinct molecular species. For a given binary system, the same model parameters were used for liquid mixtures at all compositions and for solvation free energy calculations. The valence and DATM parameters (sections V.B and V.C, respectively) were optimized separately for each molecular species and were shared across all systems containing that species. In contrast, the parameters for intermolecular multipole–multipole interactions (section V.A) were optimized separately for each system, allowing these parameters for a given molecule to differ between its neat-liquid model and models of binary systems containing that molecule. For each system, we generated an ensemble of five models with different multipole and polarization parameters but identical valence and DATM parameters.

### *A. Intermolecular multipole-multipole interactions*

We fit together all model parameters (including atomic polarizabilities used in the induction model) impacting multipole-multipole energy contributions from dynamic, static, and induced multipoles. The QM dataset included the following types of data, for multiple geometries (see section IX for more information about the QM calculations):

- dimer total interaction energy (604,273 geometries)
- multimer total interaction energy and/or its nonadditive component (4095 geometries)
- microsolvation total interaction energy (which relates, with the opposite sign, to the energy of extracting a central solute molecule from a cluster with multiple solvent molecules) and its nonadditive component (9054 geometries)
- monomer electrostatic potential map
- monomer dipole, quadrupole, and polarizability
- dimer long-range dispersion coefficients $C_6$ and $C_8$

All geometries were extracted from PIMD simulations with a preliminary force field to ensure adequate sampling of substantial intramolecular deformations arising from NQEs, which would be less frequently sampled in classical MD simulations.

Model parameters were fit with a nonlinear least-squares procedure using an analytical Jacobian matrix of first derivatives of residuals with respect to each parameter. The DATM energy term was excluded from this fitting procedure because the reference nonadditive interaction energies were computed at the MP2 level, which does not account for three-body and higher-order dispersion contributions.

For each system, the QM dataset was typically divided into training and validation sets in a 60:40 ratio, although an 80:20 split was used in some cases. We attempted 200 optimizations starting from different initializations and selected the five models with the lowest validation loss, rather than the lowest training loss, to form the model ensemble. The validation errors were typically within 20% of the corresponding training errors, suggesting that substantial overfitting did not occur. The results shown in Fig. 1 of the main text include both training and validation geometries. Ultimately, however, a more stringent test of the predictive accuracy of DES-XFF is provided by comparisons with experimental condensed-phase properties, none of which were used in fitting or selecting the models.

### *B. Intramolecular valence interactions*

Intramolecular valence models were fit in two stages. First, for each molecule, we used $10^5$ single-point QM energies computed at a relatively inexpensive MP2 level of theory to fit a complete model containing all diagonal terms up to fourth order and all cross-terms up to third order. This model was then subjected to an iterative feature-selection procedure (backward selection) in which at each iteration we removed the cross-term with the least impact on the value of the objective function. Iterations stopped when the value of the objective function exceeded a specified threshold. During the second stage, we used $10^4$ single-point QM energies computed at a very accurate PNO-LCCSD(T*)-F12b level of theory to fit the final model with the same diagonal terms and cross-terms, as determined by the backward-selection procedure described above.

### C. Intermolecular many-body dispersion interactions

We fit our many-body dispersion (MBD) models using a two-stage protocol. First, we used monomer polarizabilities and long-range three-body dispersion $C_9$ coefficients to fit a coupled fluctuating dipole model (CFDM)[5,6] for many-body dispersion. This model was then used to generate MBD energies for periodic boxes representative of the liquid state. During the second stage, we used these CFDM energies for periodic boxes as a reference to fit a three-body DATM model.

## VI. Details of MD simulations

### A. Simulation protocols

All condensed-phase simulations were performed with a modified version of the Desmond/GPU software.[7] Long-range electrostatics were handled using the *u*-series decomposition[8] ($C^2$-continuous solution with $b \approx 1.32$). To support higher-order derivatives of the transfer function required for charge-spreading of multipoles, we generalized midtown splines[9] to be $C^2$-continuous in these higher-order derivatives. We used a 7-point quadrature spline in calculations involving octupoles, and a 6-point quadrature spline in all other cases. Maximum grid cell size was set at 0.04 times the cutoff radius, which ensured ~$10^{-3}$ kcal/mol/molecule accuracy in electrostatic energy. The cutoff radius was set at 10 Å for near electrostatics and dispersion, and at 7 Å for exchange. To compute the multipolar pair interactions, we used a previously published algorithm[10] that involved rearranging terms into new intermediate quantities to obtain an efficient expression for the gradient with respect to multipole moments. This gradient was necessary because (i) all multipole moments are defined in LORAS and (ii) dynamic multipoles are explicitly dependent on the positions of neighboring atoms. The long-range dispersion correction included $C_6$ and $C_8$ contributions to both the energy and (because of the dynamic multipole model for dispersion) the atomic forces. The induced dipole degrees of freedom were handled using an in-house version of the extended Lagrangian approach,[11] which was validated against self-consistent field calculations.

In all production simulations, the unit cell size was approximately 30 × 30 × 30 Å$^3$. Trajectories were propagated with the BAOAB integrator[12] and the time step was 0.5 fs (1.0 fs in the PIMD solvation free energy simulations; Table VI.I). Temperature was controlled using the Langevin thermostat[13] with a friction coefficient of 1 ps$^{-1}$. In all NPT simulations, the pressure was maintained at 1 atm using a Monte Carlo (MC) barostat[14] applied every 48 time steps. In PIMD simulations, we used $P = 64$ ring-polymer beads, except for the ethane and ethylene simulations, which were performed with $P = 256$ beads to improve accuracy at low temperatures. Normal-mode transformation was used to diagonalize the mass matrix. We propagated the free ring-polymer evolution exactly and applied optimal friction coefficients to each bead.[15] To improve computational efficiency, we employed the ring-polymer contraction scheme[16] to evaluate the non-bonded forces on six beads (32 for ethane and ethylene), designated, for example, as $P = 64/6$. Solvation free energies were computed using an implementation of Times Square sampling[17] (TSS) for free energy perturbation calculations with a 0.1 ps TSS interval and a schedule that transforms the solute-solvent interactions into OPLS potentials[18] before turning them off.

**Table VI.I**. PIMD and classical MD setup for various types of simulations. Neat liquid, mixed liquid, and water ice simulations were performed for each of the force fields in the ensemble for each system. For the gas-phase simulations, as there is only one valence force field per system, only one simulation was needed. For all neat liquid and mixed liquid simulations, we performed either 5 or 10 replicate (i.e., repeat) simulations for each of the force fields in the ensemble. For all solvation free energy simulations (which were either 10 ns or 80 ns in length), only the top-ranked force field in the ensemble (the one with the lowest loss on the QM validation set) was used, and only one simulation was performed.

| Name | Phase | PIMD beads[a] | dt (fs) | Length[b] (ns) | Replicates | Thermostat/ Barostat | Ensemble |
|---|---|---|---|---|---|---|---|
| **Neat** | Gas | 1/1 | 0.5 | 1.3/0.3 | ≥180[c] | Langevin | NVT |
| **Neat** | Gas | 64/6 | 0.5 | 1.3/0.3 | ≥180 | Langevin | NVT |
| **Neat, Ethane/Ethylene** | Gas | 256/32 | 0.5 | 1.3/0.3 | ≥180 | Langevin | NVT |
| **Neat** | Liquid | 1/1 | 0.5 | 2.5/0.5 | 5 | Langevin/MC | NPT |
| **Neat** | Liquid | 64/6 | 0.5 | 2.5/0.5 | 5 | Langevin/MC | NPT |

[a] PIMD bead counts are shown as total/contracted; 1/1 is equivalent to running classical MD.

[b] Trajectory length format is total time/burn time; burn time is the initial equilibration segment excluded from statistical analysis.

[c] Because we turned off intermolecular interactions, the N monomers effectively form N replicate simulations (where N was ≥180).

| | | | | | | | |
|---|---|---|---|---|---|---|---|
| **Neat, Ethane/Ethylene** | Liquid | 256/32 | 0.5 | 2.5/0.5 | 5 | Langevin/MC | NPT |
| **Neat, Water Ice** | Solid | 1/1 | 0.5 | 5.0/0.5 | 1 | Langevin/MC | NPT |
| **Neat, Water Ice** | Solid | 64/6 | 0.5 | 5.0/0.5 | 1 | Langevin/MC | NPT |
| **Neat, Water Long** | Liquid | 1/1 | 0.5 | 5.0/1.0 | 10 | Langevin/MC | NPT |
| **Neat, Water Long** | Liquid | 64/6 | 0.5 | 5.0/1.0 | 10 | Langevin/MC | NPT |
| **Mixed** | Liquid | 1/1 | 0.5 | 2.5/0.5 | 5 | Langevin/MC | NPT |
| **Mixed** | Liquid | 64/6 | 0.5 | 2.5/0.5 | 5 | Langevin/MC | NPT |
| **Solvation free energy** | Liquid | 1/1 | 0.5 | 10.0/0[d] | 1 | Langevin/MC | NPT |
| **Solvation free energy** | Liquid | 64/6 | 1.0 | 80.0/0 | 1 | Mixed | NPT |

[d] For the solvation free energy simulations, which were all performed using multi-replica TSS with four replicas, the burn time of 0 indicates that sampling begins at the first time step (i.e., there is effectively no burn time).

### *B. Computational performance*

We benchmarked the computational performance of DES-XFF for classical MD and PIMD simulations of periodic boxes of liquid water of varying sizes on NVIDIA A100, H200, and RTX PRO 6000 Blackwell GPUs (Fig. VI.1). On a single RTX PRO 6000 Blackwell GPU, classical MD simulation throughput ranged from approximately 56 ns/day for a system of 1650 atoms (550 water molecules; box length of 25.9 Å) to 2.7 ns/day for a system of 100,002 atoms (33,334 water molecules; box length of 102.2 Å), with corresponding PIMD throughputs of 22 and 0.37 ns/day, respectively. Despite the use of 64 ring-polymer beads, PIMD simulations on the RTX PRO 6000 Blackwell GPU were only approximately 2.5–7.3 times as computationally expensive as classical MD simulations, owing in part to the use of ring-polymer contraction. For most system sizes, simulation throughput on the RTX PRO 6000 Blackwell GPU was approximately 2.5 times that on the A100 GPU, which was used for most of the production simulations in this study.

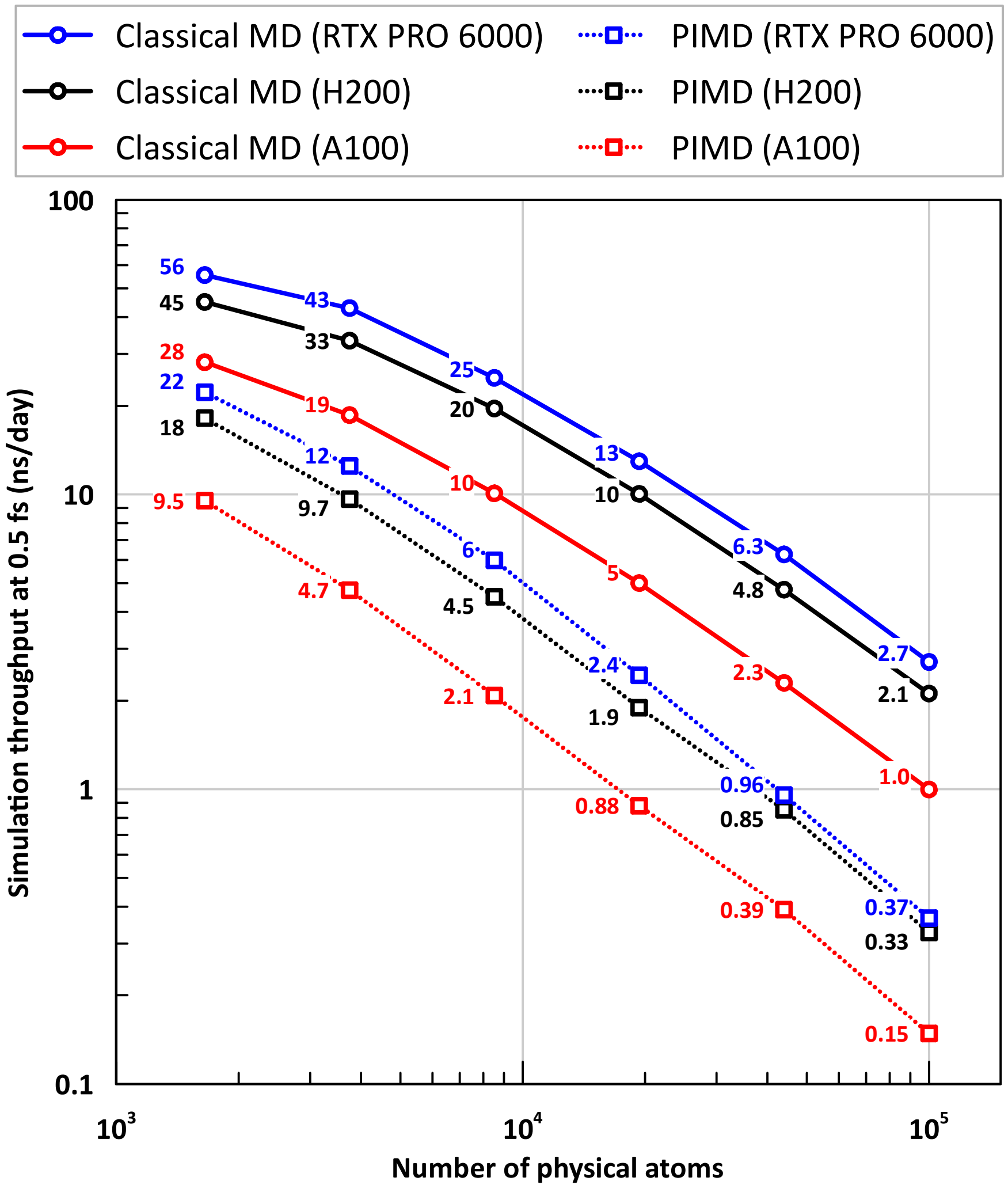

**Fig. VI.1.** Computational performance of DES-XFF for classical and PIMD simulations. Simulation throughput as a function of system size for periodic boxes of pure water using a single NVIDIA RTX PRO 6000 Blackwell (blue), H200 (black), or A100 (red) GPU. Solid lines with circles denote classical MD and dotted lines with squares denote PIMD. All simulations included MBD interactions and used an integration time step of 0.5 fs. PIMD simulations employed 64 ring-polymer beads, contracted to 6 beads for the evaluation of intermolecular interactions. The number of physical atoms excludes the additional replicas used in PIMD. Numerical labels indicate simulation throughputs in ns/day.

## VII. Condensed-phase results

This section contains additional results for condensed-phase properties of the systems studied.

### *A. Water properties*

Water properties are summarized in Table VII.I, and radial distribution functions are shown in Fig. VII.1.

**Table VII.I.** Comparison of DES-XFF predictions for neat water properties with experiment. DES-XFF values were obtained using the relevant settings given in Table VI.I (specifically, for "Neat, Water Long" and "Neat, Water Ice"). Uncertainties correspond to the 95% confidence interval.

| Property | Units | Temperature | DES-XFF | Experiment | DES-XFF error |
|---|---|---|---|---|---|
| **Density $H_2O$** | g/cm$^3$ | 25 °C | 0.996 ± 0.001 | 0.9970 | 0.1% |
| **Density $D_2O$** | g/cm$^3$ | 25 °C | 1.102 ± 0.001 | 1.1045 | 0.2% |
| **Density $H_2O$ classical** | g/cm$^3$ | 25 °C | 0.983 ± 0.001 | | |
| **Ice[a] density $H_2O$** | g/cm$^3$ | 0 °C | 0.901 ± 0.001 | 0.9167 | 1.7% |
| **Ice density $D_2O$** | g/cm$^3$ | 4 °C | 1.002 ± 0.001 | 1.0175 | 1.5% |
| **Ice density $H_2O$ classical** | g/cm$^3$ | 0 °C | 0.896 ± 0.001 | | |
| **TMD[b] $H_2O$** | °C | | 2 ± 2 | 4 | |
| **TMD $D_2O$** | °C | | 13 ± 2 | 11 | |
| **TMD $H_2O$ classical** | °C | | 18 ± 2 | | |
| **$\Delta H_{vap}$ $H_2O$** | kcal/mol | 25 °C | 10.18 ± 0.02 | 10.51 | 3% |
| **$\Delta H_{vap}$ $D_2O$** | kcal/mol | 25 °C | 10.58 ± 0.02 | 10.84[c] | 2% |
| **$\Delta H_{vap}$ $D_2O$** | kcal/mol | 50 °C[d] | 10.30 ± 0.02 | 10.53 | 2% |

[a] All ice simulations were performed with the unit cell containing 768 water molecules, based on the system described in Ref. 19.

[b] TMD stands for temperature of maximum density.

[c] Using the experimental value for the isotope effect on $\Delta H_{vap}$ of 0.33 kcal/mol from Ref. 20.

[d] Experimental $\Delta H_{vap}$ values for $D_2O$ are from Ref. 21; data are applicable only above 50 °C.

| | | | | | |
|---|---|---|---|---|---|
| **Δ$H_{vap}$ $H_2O$ classical** | kcal/mol | 25 °C | 10.83 ± 0.01 | | |
| **Δ$H_{fus}$ $H_2O$** | kcal/mol | 0 °C | 1.53 ± 0.02 | 1.44 | 6% |
| **Δ$H_{fus}$ $D_2O$** | kcal/mol | 4 °C | 1.61 ± 0.02 | 1.47–1.51[e] | 7% |
| **Δ$H_{fus}$ $H_2O$ classical** | kcal/mol | 0 °C | 1.64 ± 0.01 | | |
| **Δ$G_{solv}$ $H_2O$** | kcal/mol | 25 °C | −6.17 ± 0.11 | −6.3 | 2% |
| **Δ$G_{solv}$ $H_2O$ classical** | kcal/mol | 25 °C | −6.29 ± 0.05 | | |
| **Heat capacity $C_p$ $H_2O$** | cal/mol/K | 25 °C | 17.4[f] ± 0.2 | 18.0 | 3% |
| **Compressibility $H_2O$** | 1/GPa | 25 °C | 0.48 ± 0.04 | 0.45 | 7% |

[e] Using experimental values from Ref. 22 and references therein.

[f] Isobaric heat capacity at 1 atm for DES-XFF is estimated from the slope of liquid $H_2O$ enthalpy as a function of temperature, giving 18.3 cal/mol/K. This value is then corrected for the spurious intramolecular heat capacity of $H_2O$ vapor, 0.9 cal/mol/K, which differs from the correct value of near 0.0 cal/mol/K due to remaining discretization artifacts in PIMD with $P$ = 64 beads.

| | | | | | |
|---|---|---|---|---|---|
| **Diffusion coefficient $H_2O$** | Å$^2$/ps | 25 °C | 0.280[g] ± 0.014 | 0.230 | 22% |
| **Diffusion coefficient $H_2O$ classical** | Å$^2$/ps | 25 °C | 0.253[h] ± 0.013 | | |
| **Diffusion coefficient $D_2O$** | Å$^2$/ps | 25 °C | 0.211[i] ± 0.012 | 0.187 | 13% |
| **Diffusion coefficient $D_2O$ classical** | Å$^2$/ps | 25 °C | 0.219 ± 0.012 | | |
| **Second virial $B_2$ $H_2O$** | cm$^3$/mol | 400 K | −340.1[j] ± 0.1 | −351 ± 14 | 3% |

[g] DES-XFF predicts a 10.7% increase in the $H_2O$ diffusion coefficient due to NQEs. This result was obtained using the RPMD procedure detailed in Refs. 23 and 24.

[h] Classical diffusion coefficient was estimated from the slope of the mean-square displacement and corrected for finite-size effects, as recommended in Ref. 25.

[i] RPMD calculations with DES-XFF predict a 3.7% decrease of the $D_2O$ diffusion coefficient due to NQEs, a difference that is within the statistical uncertainty.

[j] The second virial coefficient was computed using flexible monomers, accounting for both intramolecular and intermolecular NQEs.

| | | | | | |
|---|---|---|---|---|---|
| **Second virial $B_2$ $H_2O$** | cm$^3$/mol | 700 K | −64.40 ± 0.01 | −65.9 ± 2.6 | 2% |

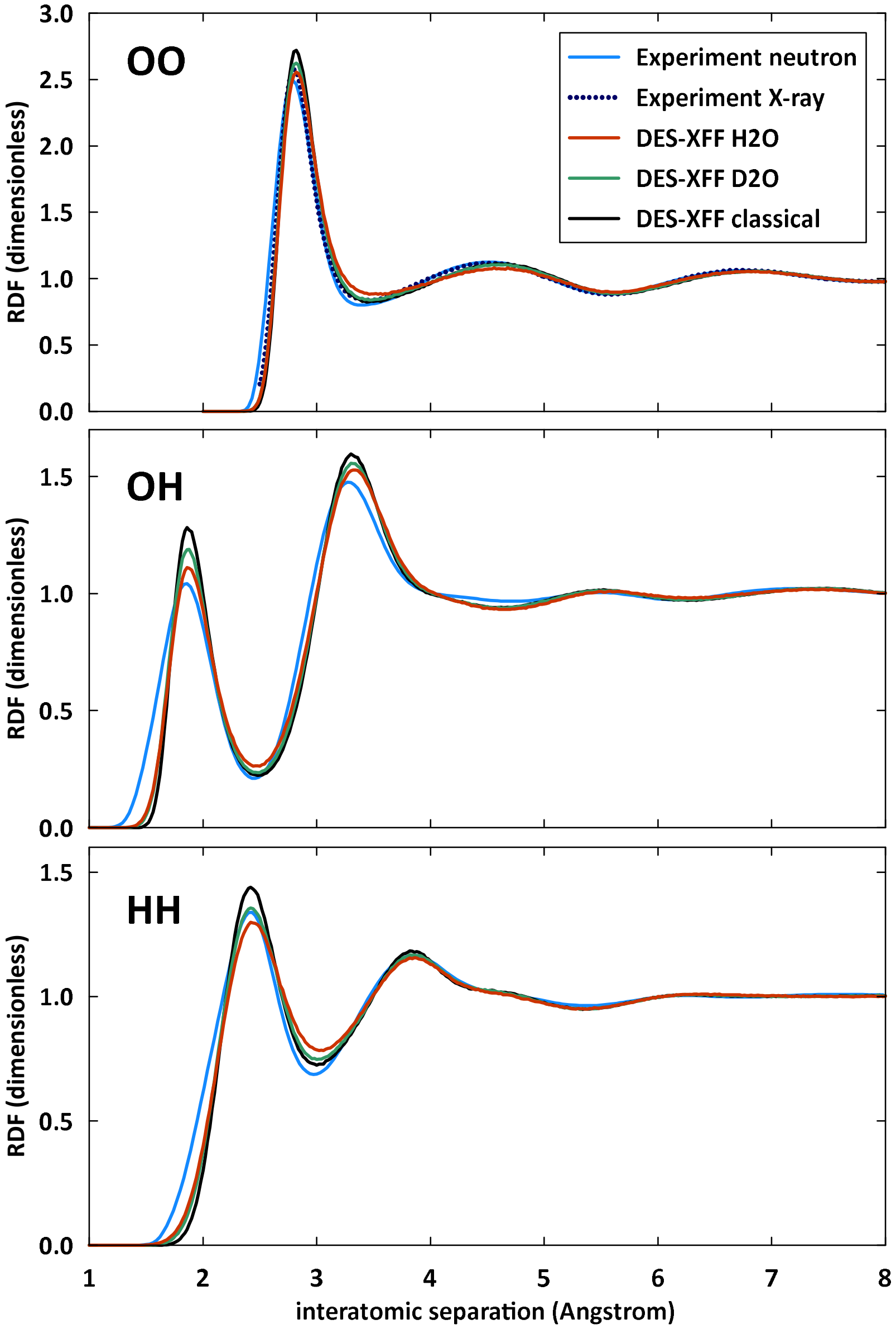

**Fig. VII.1.** Radial distribution functions for liquid water at 25 °C and 1 atm. DES-XFF predictions from PIMD simulations of $H_2O$ and $D_2O$ and classical MD simulations are compared with experimental data. All simulations used the top-ranked DES-XFF model (i.e., the model

with the lowest loss on the QM validation set). Neutron diffraction experimental data is from Ref. 26; X-ray experimental data is from Ref. 27.

### *B. Excess properties of mixtures*

Mixture properties are presented in Fig. VII.2.

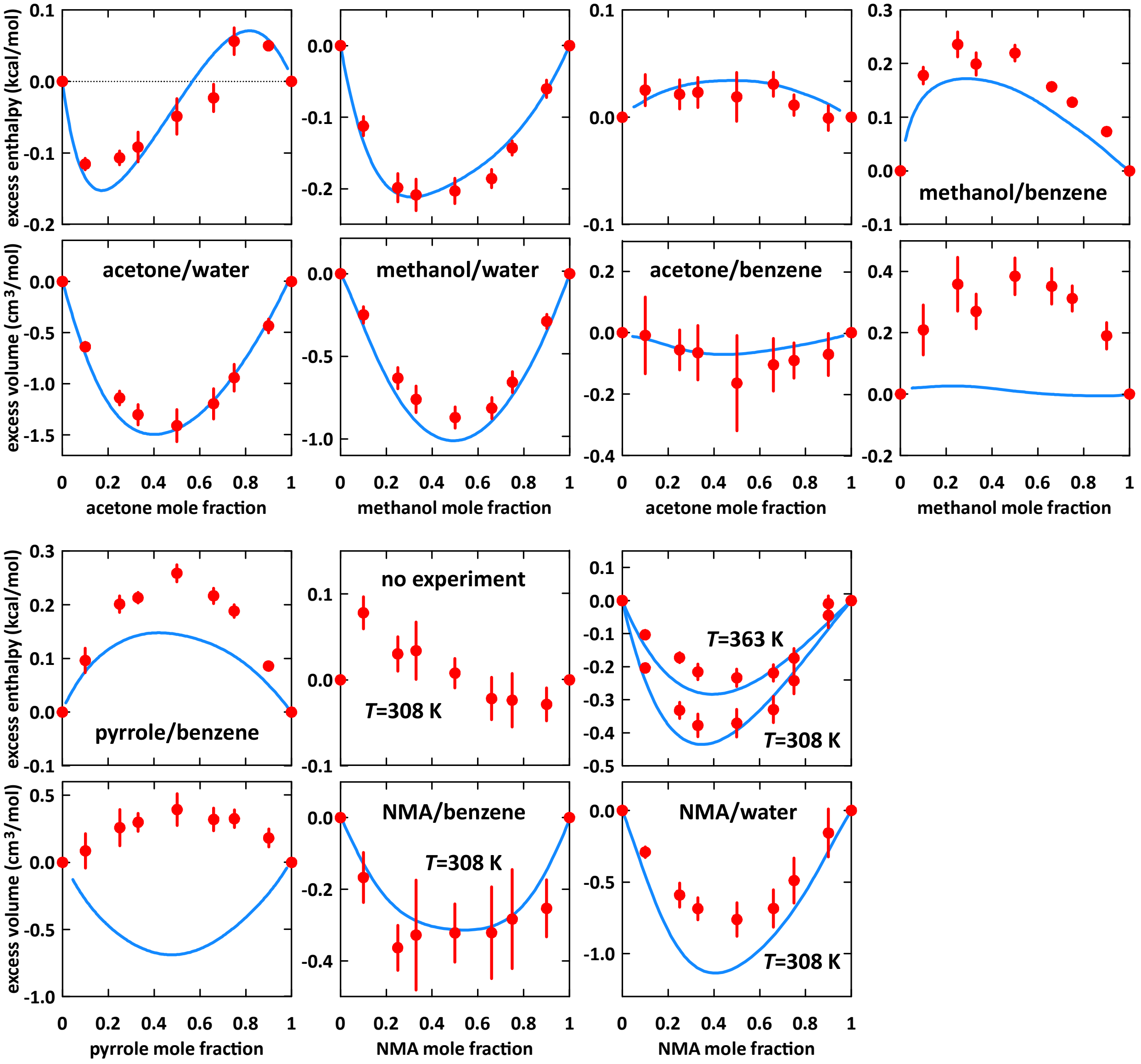

**Fig. VII.2.** Comparison of DES-XFF predictions from PIMD simulations (red circles) with experimental values (blue lines) for the excess enthalpy (upper panel of each pair) and excess volume (lower panel) of seven binary liquid mixtures. All results are at 298 K unless otherwise stated. DES-XFF predictions are averaged over the five corresponding models, and the error bars represent 95% confidence intervals.

### *C. Effect of QM data accuracy*

The development of DES-XFF relied on a large amount of coupled-cluster and MP2-level QM data. Even with the PNO approximations and composite basis set approach that we used in generating the coupled-cluster-level data (see section IX), however, PNO-LCCSD(T*)-F12b/aVXZ-F12 + MP2/CBS dimer interaction calculations are too expensive to run at the scale needed to extend DES-XFF to cover the full breadth of chemistry represented in drug-like molecules. An important question is thus what would happen if less accurate but faster QM methods were used instead. QM data generated using less costly, and slightly less accurate, DFT-based methods[28] has become popular in fitting ML-based force fields,[29–32] and approaches such as ωB97M-V[33] can reproduce gold-standard dimer interaction energies to within ~0.1 kcal/mol on standard test sets such as S66.[34,35] The errors of DFT-based methods, however, are neither identically nor independently distributed with respect to chemistry or configuration,[36] and tend to grow as the sampled structures deviate further from the QM optimum geometry. These methods can also struggle to predict many-body contributions.[37] As a test of the impact of these errors, for a subset of our systems we reran the dimer (and in some cases multimer) QM calculations used in our fits using ωB97M-V/def2-TZVPPD,[33] refit the models to this data, and used the resulting force fields to perform MD simulations. These force fields let us quantify the impact of the use of cheaper QM data on condensed-phase properties, as shown in Figs. VII.3 and VII.4.

Replacing the coupled-cluster-level dimer interaction energies with ωB97M-V dimer interaction energies led to significant degradation in condensed-phase performance. The difference was most prominent for benzene, which moved far outside ±2% and ±0.5 kcal/mol errors for density and $\Delta H_{vap}$ predictions, respectively. Interestingly, removing MBD contributions by disabling DATM during our simulations provided a cancellation of errors that brought the ωB97M-V/MP2 results into closer agreement with experiment, likely because ωB97M-V underestimates the strength of dimer interactions relative to coupled-cluster calculations (Table X.III). This cancellation of errors, however, depends on the choice of QM reference data for both the two-body and nonadditive interaction energies.

For benzene, we also investigated a model in which both the dimer and nonadditive interaction energies were computed using ωB97M-V. In contrast to the model trained on ωB97M-V dimer energies and MP2 nonadditive energies, this model substantially overestimated both liquid density and vaporization enthalpy, irrespective of whether MBD was included (red markers in Fig. VII.3). With DATM included, the liquid density was overestimated by approximately 5%–7% and $\Delta H_{vap}$ by 0.3–0.5 kcal/mol; without DATM, the corresponding overestimations were approximately 13%–14% and 1.5–1.7 kcal/mol. These results illustrate the sensitivity of condensed-phase predictions to the accuracy of the nonadditive QM reference energies and suggest that the favorable cancellation of errors observed for the ωB97M-V/MP2 model without DATM cannot be expected to occur generally. This finding is particularly relevant to the development of machine-learned force fields trained exclusively on DFT interaction energies of molecular clusters.

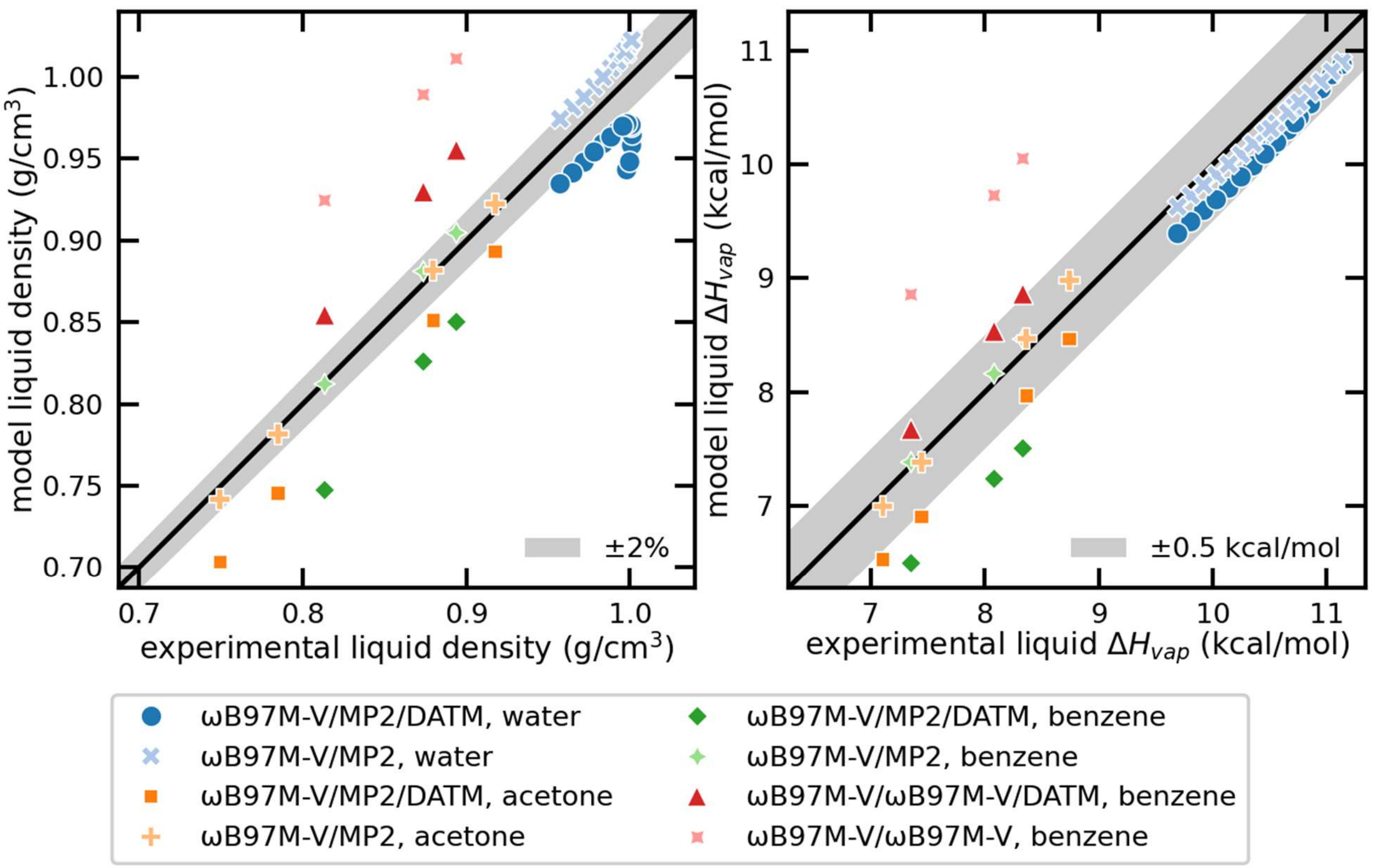

**Fig. VII.3.** Condensed-phase properties of water (blue markers), acetone (orange markers), and benzene (green and red markers) predicted by models fit to dimer interaction energies calculated using ωB97M-V/def2-TZVPPD rather than PNO-LCCSD(T*)-F12b/aVXZ-F12 + MP2/CBS. In the case of benzene, we also recalculated the nonadditive energies using ωB97M-V/def2-TZVPPD (red markers) instead of MP2. In all cases, we performed simulations both with (dark markers) and without (pale markers) DATM enabled. The results shown represent PIMD calculations at the same temperatures for each system as the temperatures used in Fig. 3 of the main text.

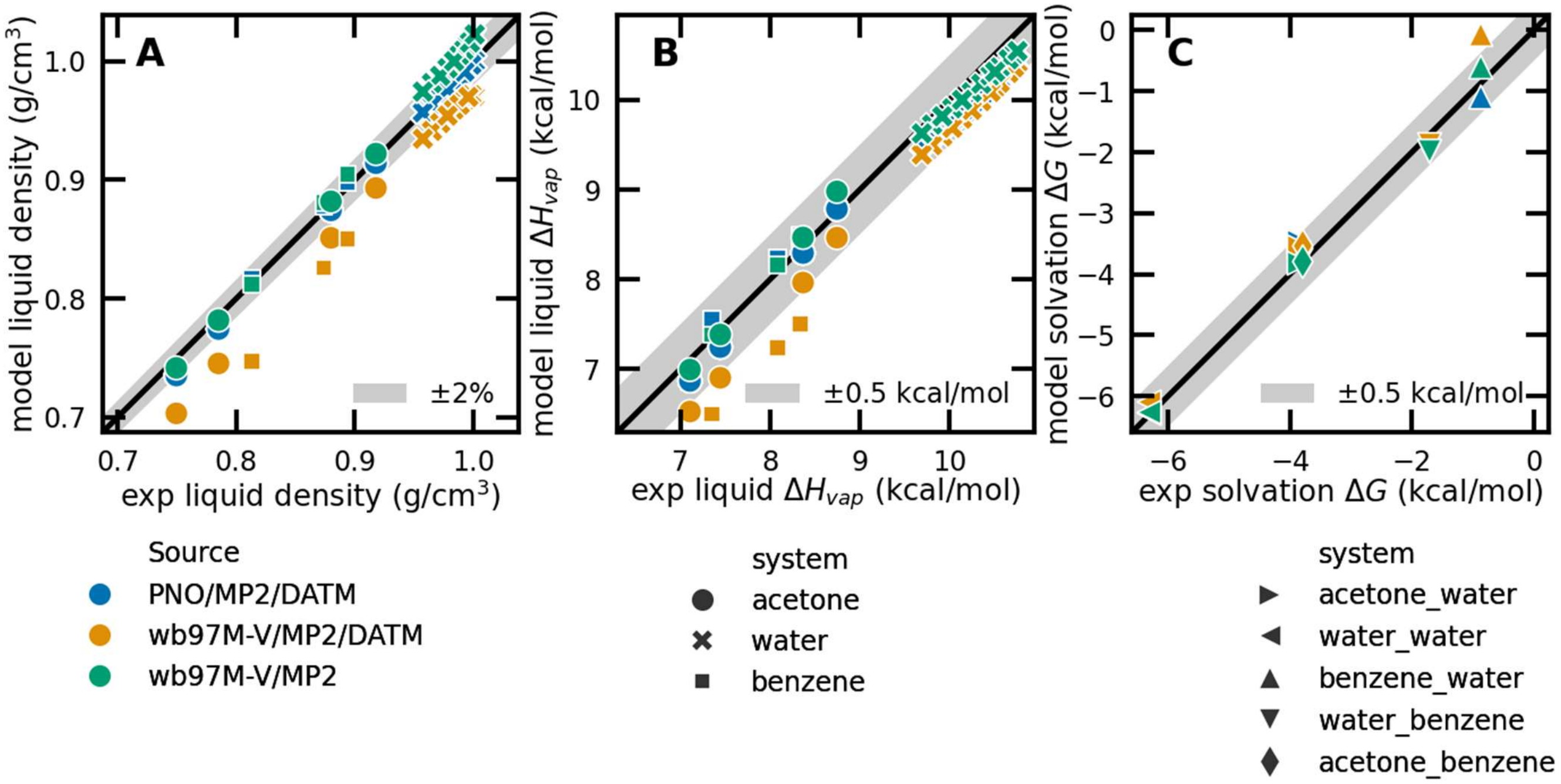

**Fig. VII.4.** Effect of QM data accuracy on predictions of condensed-phase properties. **(A)** Comparison of predictions of three sets of models with experimental liquid densities at various temperatures between the melting and boiling points of water, acetone, and benzene. PNO/MP2/DATM corresponds to DES-XFF, ωB97M-V/MP2/DATM corresponds to a set of models with the same DES-XFF functional form but with the dimer data computed using ωB97M-V/def2-TZVPPD, and ωB97M-V/MP2 is the same but simulated without MBD included. **(B)** Same as (A) but for vaporization enthalpy $\Delta H_{vap}$. **(C)** Comparison of predictions, computed using classical simulations, of three sets of models with experimental solvation free energies ($\Delta G_{solv}$). In (C), the system labels have the form solute_solvent.

### *D. Comparison of average contributions of NQEs and MBD*

Table VII.II summarizes the average contributions of NQEs and MBD to the properties discussed in the main text.

**Table VII.II.** Average contributions of NQEs and MBD to condensed-phase properties predicted by DES-XFF. All values are signed averages over the systems and temperatures specified in each row. For liquid density and vaporization enthalpy, averages are reported for all 12 liquids (including water) over all studied temperatures, for the 11 organic liquids over all studied temperatures, and for 9 of these organic liquids over the 273–373 K range (excluding ethane and ethylene, which boil below 273 K). NQE contributions are calculated as differences between PIMD and classical MD simulations, whereas MBD contributions are calculated as differences between PIMD simulations with and without MBD.

| Property | Average NQE correction | Average MBD correction |
|---|---|---|
| Liquid density (11 organic liquids, all temperatures studied) | −0.9% | −4.6% |
| Liquid density (9 organic liquids, 273–373 K) | −0.6% | −4.9% |
| Liquid density (12 liquids, including water, all temperatures studied) | −0.7% | −4.6% |
| $\Delta H_{vap}$ (11 organic liquids, all temperatures studied) | −0.08 kcal/mol | −0.61 kcal/mol |
| $\Delta H_{vap}$ (9 organic liquids, 273–373 K) | −0.05 kcal/mol | −0.66 kcal/mol |
| $\Delta H_{vap}$ (12 liquids, including water, all temperatures studied) | −0.13 kcal/mol | −0.57 kcal/mol |
| $\Delta G_{solv}$ (12 solutes in water, room temperature) | +0.17 kcal/mol | +0.30 kcal/mol |
| $\Delta G_{solv}$ (6 solutes in benzene, room temperature) | −0.04 kcal/mol | +0.30 kcal/mol |

## VIII. Experimental data

Experimental values for neat liquid densities and vaporization enthalpies, $\Delta H_{vap}$, were derived, as described below, using data from two sources:

i) The Design Institute for Physical Properties 801 database (DIPPR),[38] using the public/licensed version downloaded in March 2018, from which we used only the "A-quality" values, and

ii) NIST ThermoData Engine (TDE),[39] from which we used densities with reported uncertainty less than 25 kg/m$^3$. (We additionally removed two outliers from the density set for dimethylamine.)

We found the $\Delta H_{vap}$ data in DIPPR to be much more consistent as a function of temperature than the $\Delta H_{vap}$ data in TDE. This prompted us to use only DIPPR as a source of $\Delta H_{vap}$ data, except in the case of methanol, for which we used both sources because TDE provided data at lower temperatures. Even using the TDE set, the $\Delta H_{vap}$ data for methanol is not available for temperatures below 273 K, which renders extrapolation to the vicinity of methanol's freezing point (175 K) somewhat unreliable (though it is still used for our comparisons). For imidazole, $\Delta H_{vap}$ data was not available in either database, and so we estimated the $\Delta H_{vap}$ value at the melting point (362 K) from the experimental enthalpies of sublimation and fusion, as explained below.

To facilitate comparisons between model predictions and experiment at arbitrary temperatures (from the melting point to the boiling point), experimental data was approximated by second-degree polynomials using a least-squares fitting procedure with uniform weights. Table VIII.I provides the quality of fit for each system.

**Table VIII.I.** Quality of fits to experimental data.

| Liquid | Melting Point (K) | Boiling Point (K) | Mean uncertainty of DIPPR $\Delta H_{vap}$ values (kcal/mol) | Fitting RMSD $\Delta H_{vap}$ (kcal/mol) | Fitting RMSD density (kg/m$^3$) |
|---|---|---|---|---|---|
| Water | 273.1 | 373.1 | 0.10 | 0.001 | 0.36 |
| NMA | 301.1 | 478.1 | 1.30 | 0.007 | 2.24 |
| Phenol | 313.6 | 454.8 | 0.37 | 0.002 | 2.04 |
| Acetone | 178.2 | 329.1 | 0.24 | 0.024 | 1.74 |
| Benzene | 278.6 | 353.2 | 0.08 | 0.008 | 0.64 |
| Dimethylamine | 180.1 | 280.1 | 0.21 | 0.011 | 0.54 |
| Dimethyl sulfide | 175.0 | 310.5 | 0.19 | 0.014 | 1.07 |
| Ethane | 90.4 | 184.2 | 0.04 | 0.006 | 1.15 |
| Ethylene | 104.0 | 169.5 | 0.04 | 0.010 | 0.82 |
| Methanol | 175.4 | 337.8 | 0.26 | 0.036 | 0.73 |
| Pyrrole | 250.2 | 402.1 | 0.30 | 0.056 | 0.36 |
| Imidazole | 362.0 | 529.1 | N/A | N/A | 4.03 |

### A. *Derivation of imidazole* $\Delta H_{vap}$ *value at the melting point*

Because $\Delta H_{vap}$ data for imidazole were absent from both DIPPR and TDE, we estimated the $\Delta H_{vap}$ value at the melting point (362 K) using the experimental enthalpies of sublimation ($\Delta H_{subl}$) and fusion ($\Delta H_{fus}$).

We used the standard relationship between $\Delta H_{vap}$, $\Delta H_{subl}$, and $\Delta H_{fus}$,

$$\Delta H_{vap}(T) = \Delta H_{subl}(T) - \Delta H_{fus}(T) \ ,$$

where $\Delta H_{subl}(T)$ at an arbitrary temperature $T$ can be estimated from the standard value $\Delta H_{subl}$(298 K) and the crystal's isobaric heat capacity $C_p$ at room temperature,[40]

$$\Delta H_{subl}(T_1) - \Delta H_{subl}(T_2) = -\left[0.75 + 0.15 C_p(298\text{K})\right] \times (T_1 - T_2)/1000 \ .$$

Combining the two equations and substituting $\Delta H_{subl}$(298 K) = 79.5 kJ/mol (average of the three most recent measurements reported in Ref. 40), $\Delta H_{fus}$(362 K) = 12.8 kJ/mol,[40] and $C_p$(298 K) = 89 J/(mol K) (average of the three values from NIST[41]) yields $\Delta H_{vap}$(362 K) = 65.8 kJ/mol = 15.7 kcal/mol.

## IX. Quantum mechanical calculations

Several types of quantum mechanical (QM) calculations were used to produce the data presented in this paper, with the method selection for different cases depending on the kind of system and the properties sought. As our datasets were developed over a long period of time, there is some variation in the versions of the QM software employed, the precise settings used, and some of the post-processing steps (such as the choice of counterpoise (CP) correction scheme).

Selecting the level of QM theory involved a trade-off between accuracy and speed, and for different kinds of molecular systems we chose different techniques. Overall, the molecular systems can be divided into four categories: monomers, dimers, multimers, and microsolvation clusters. Monomer calculations measure deformation energies of a molecule, dimer calculations measure the non-covalent interactions of two molecules, and multimer and microsolvation cluster calculations include a component that measures many-body interactions. Monomer and dimer calculations were performed at a coupled-cluster level of theory, whereas multimer and microsolvation calculations were performed using a composite approach that combines a two-body energy computed at the coupled-cluster level of theory with a many-body energy computed at the MP2 level of theory.[42] In addition, we performed dimer (and some many-body) calculations using a high-quality, dispersion-inclusive density functional (ωB97M-V) to assess the impact of using less expensive QM data.

### *A. General settings*

HF, MP2, CCSD(T)[43–45] and PNO-LCCSD(T*)-F12b[46–48] calculations were performed using various versions of the Molpro[49–51] software package. Density fitting[52,53] was used for all HF and MP2 calculations except in the HF calculations preceding a CCSD(T) calculation. All PNO-LCCSD(T*)-F12b calculations used the DOMOPT=TIGHT setting available in Molpro,[54,55] and used a DF-MP2-F12 calculation as the MP2-level reference. (That is, δPNO-LCCSD(T*)-F12b = PNO-LCCSD(T*)-F12b – DF-MP2-F12. Upon further analysis, we found that using the direct output of the PNO-LCCSD(T*)-F12b/haVXZ-F12 calculations (that is, HF + CABS + PNO-LMP2-F12 + PNO-LCCSD + (T)) provided slightly better agreement with gold-standard calculations, but the differences were below our fitting errors.) Triples scaling for PNO-LCCSD(T*)-F12b was computed in a size-consistent fashion using the MP2-F12/MP2 ratio from the largest subsystem relevant to a calculation as a scale factor.[56] Note that some of the settings and defaults of the PNO program in Molpro changed between the 2019 and the 2020+ versions of the code; these differences slightly impacted the calculated interaction energies.[57]

ωB97M-V[33] and SAPT0[58–61] calculations were performed using various versions of Psi4.[62] DFT calculations used the SBECKE atomic weighting scheme, the EM radial grid scheme, and the robust grid-point pruning scheme. SAPT0 calculations were performed using density fitting and a monomer SAPT basis.

For most calculations, we used Dunning correlation-consistent basis sets (aug-cc-pVXZ,[63] abbreviated as aVXZ), whereas for coupled-cluster calculations with approximate F12 corrections we used newer F12 basis sets (aug-cc-pVXZ-F12,[64] abbreviated as aVXZ-F12). Occasionally, we needed to remove the diffuse functions on hydrogens, either for calculation stability or to reduce calculation costs further—these basis sets are indicated as haVXZ(-F12). DFT calculations used the Turbomole def2 basis set def2-TZVPPD.[65] The default matching auxiliary basis sets were used in all calculations in all software packages.

Complete-basis-set corrections for MP2 calculations were computed using a two-point Helgaker extrapolation scheme[66] (using an exponent of three). Basis set superposition error was accounted for by Boys–Bernardi CP corrections,[67] with full-CP used for HF and MP2 calculations, and full- or half-CP[68] used for coupled-cluster corrections. Frozen core orbital definitions freezing all inner shell atomic orbitals were used in all cases, corresponding to CORE,LARGE in Molpro and freeze_core=True in Psi4.

### B. Monomer calculations

Monomer geometries were sampled uniformly from PIMD simulations performed using a preliminary force field. The use of PIMD ensured adequate sampling of substantial intramolecular deformations arising from NQEs, which would be less frequently sampled in classical MD simulations. The simulations were performed at 298 K if the substance is liquid at that temperature, or otherwise at its melting point or boiling point, whichever is closer to 298 K.

#### 1. Monomer deformation energies

Monomer deformation energies were computed using a composite scheme common to "gold standard" datasets, utilizing HF, MP2, and PNO-LCCSD(T*)-F12b calculations:

$$E = HF_{aVQZ} + \Delta MP2_{aV[TQ]Z} + \delta PNO\text{-}LCCSD(T^{*})\text{-}F12b_{aVXZ\text{-}F12}$$

The basis set for the PNO-level calculation was chosen according to the size of the monomer.

### *2. Monomer electronic properties*

Both static and dynamic monomer polarizabilities at imaginary frequencies were computed using time-dependent density functional theory (TD-DFT) in Molpro, using the PBE0[69] functional and an aVTZ basis. Dimer and trimer dispersion coefficients (which were required to construct the CFDM model) were calculated by integrating, respectively, the double and triple products of the monomer polarizabilities.

All other monomer properties were calculated based on relaxed HF and MP2 densities, with calculations performed in an aVQZ basis. Monomer polarizabilities were calculated from numerical gradients of the multipole moments in small applied electric fields. Monomer electrostatic potentials, gradients, and multipole moments were computed from the relaxed HF and MP2 densities at grid points sampled around the nuclei in a molecule.

## *C. Dimer calculations*

Dimers were sampled starting from liquid simulations using an in-house force field, performed either at 298 K or at the molecule's melting or boiling point if the system was not liquid at 298 K. Collected dimers were then clustered and re-sampled from clusters to improve dataset diversity.[70] Finally, each unique dimer configuration was "scanned" by increasing/decreasing the center-of-mass distance between the two monomers by successive increments of 0.1 Å. This final dataset was evaluated at a SAPT0 level of theory, and the resulting interaction energies were used to sample configurations in an energetically relevant region. Dimers from multimers and dimers from microsolvation clusters were also included in our dimer datasets.

The dimer interaction energies were evaluated using the composite scheme described in section IX.B. The basis set at the PNO level was again chosen according to the size of the dimer system.

Interaction energies are defined as follows:

$$\Delta E_{ij} = E_{ij} - E_i - E_j ,$$

where the monomer calculations were performed both in the dimer basis and in the monomer basis, such that any CP correction scheme could be used. For the HF/aVQZ and MP2/aV[T/Q]Z portions of the interaction energy, a full-CP correction scheme was always used. For the post-MP2 portion of the interaction energy, the CP correction scheme was chosen according to the chemical system in question. Further, different CP schemes were sometimes used for the PNO-LCCSD and (T) components. When CCSD(T) was used, the correlation contribution was always computed using full-CP. For fits utilizing ωB97M-V/def2-TZVPPD, the full-CP DFT energy was used as the interaction energy—no composite scheme was used (no HF or MP2).

### *D. Many-body (beyond dimer) calculations*

Two types of many-body (beyond dimer) systems were considered: multimers and microsolvation clusters.

#### *1. Multimers*

Multimers representing clusters of liquid molecules of the same type were sampled uniformly from liquid simulations using an in-house force field run under the conditions described in section IX.B. Clusters were extracted by picking a molecule to be the center, and then extracting its nearest-neighbor molecules. These systems were assessed for their binding energy—calculated as the difference between the total energy of the system, $E_{\text{mult}}$, and the sum of the energies of the (non-interacting) monomers—and its two-body and many-body contributions. The many-body energies were assessed at the MP2 level of theory, with a basis set size chosen according to the multimer size. The multimer many-body energy is defined as follows:

$$E_{MB} = E_{mult} - \sum_{i>j} \Delta E_{ij} - \sum_{i} E_i = E_{mult} - \sum_{i>j} E_{ij} + (N-2)\sum_{i} E_i$$
$$\Delta E_{ij} = E_{ij} - E_i - E_j,$$

where *N* is the number of molecules. The subsystem calculations were all performed in the basis of the full multimer, thus ensuring proper CP contributions were present in the nonadditive energies. Two-body (dimer) energies were then recomputed in the dimer basis, utilizing the same composite scheme described in section IX.B and with full-CP corrections, as described in section IX.C. The two-body and total multimer energies are then defined as follows:

$$E_{\text{2-body}} = \sum_{i>j} (E_{ij} - E_i - E_j)$$

$$E_{\text{Total}} = E_{\text{MB}} + E_{\text{2-body}}$$

### 2. *Microsolvation clusters*

Microsolvation clusters representing a solute molecule dissolved in a (possibly heterogeneous) solvent were sampled uniformly from liquid simulations using an in-house force field run either at 298 K, or at the solvent's melting or boiling point if the solvent is not liquid at 298 K. Clusters were formed by extracting the solute together with a number of its nearest solvent molecules. For each microsolvation cluster, we computed the interaction energy of the solute with the surrounding solvent molecules (defined as the total energy of the cluster minus the energies of the isolated solute and the solvent cluster), decomposed into two-body and nonadditive many-body contributions. The many-body energies were assessed at the MP2 level of theory, using an aVDZ basis for the full system (all solvents + solute), and using an aVTZ basis for the reduced system (the $N$ most significant solvent molecules + solute), with $N$ being dependent on system size. The nonadditive component of the microsolvation energy is then defined as follows:

$$E_{\mu\text{Solv}} = E_{\text{all}} - E_{\text{solvent}} - E_{\text{solute}} - \sum_{i \in \text{solvent}} \left(E_{i,\text{solute}} - E_i - E_{\text{solute}}\right)$$

$$\underset{aV[D-T]Z}{E_{\mu\text{Solv}}} = \underset{DZ}{\underset{\text{full}}{E_{\mu\text{Solv}}}} + \left( \underset{TZ}{\underset{\text{reduced}}{E_{\mu\text{Solv}}}} - \underset{DZ}{\underset{\text{reduced}}{E_{\mu\text{Solv}}}} \right)$$

The energy of the entire system is given by $E_{\text{all}}$, with the total energy of the solvent portion given by $E_{\text{solvent}}$ and the total energy of the solute portion given by $E_{\text{solute}}$. The $N$ most significant solvent molecules were chosen as those with the largest magnitude of the electric field (computed using DF-MP2/aVDZ) produced by the solute molecule at the solvent molecule's center of mass. The two-body (dimer) energies were then recomputed in the dimer basis, utilizing the same composite scheme described in section IX.B and with full-CP corrections, as described in section IX.C. The two-body and total microsolvation energies are then defined as follows:

$$E_{\text{2-body}} = \sum_{i \in \text{solvent}} \left(E_{i,\text{solute}} - E_i - E_{\text{solute}}\right)$$

$$E_{\text{total}} = E_{\mu\text{Solv}} + E_{\text{2-body}}$$

## X. QM benchmarking

We used CCSD(T)/aVTZ to generate our training data for water. For all other chemical systems, however, a canonical CCSD(T) approach, either in a large basis set or using explicitly correlated F12 corrections to make up for a smaller basis, would have been prohibitively costly. We thus used PNO-LCCSD(T*)-F12b for evaluating the post-MP2 correlation contributions to our (non-water) dimer energies, as it is much cheaper than canonical CCSD(T) approaches and has demonstrated impressive accuracy[56,71] on commonly used benchmark datasets such as S22[34] and S66.[35] Because we chose our QM method according to evaluations on such datasets, however, these datasets effectively become the *validation* set for our method selection. Retrospectively, we have used a set of our own geometries, focused specifically on the chemical systems described by our family of force fields, to evaluate how much QM error our method choice introduces by comparing to reference calculations we performed on this set.

### *A. Electrum-standard calculations*

We performed reference calculations at a CCSD(T*)-F12 level. The calculations were performed in a composite, three-step fashion similar to the so-called "14K-GOLD" CCSD(T*)-F12 level calculations presented in Nagy et al.,[72] which reproduce canonical coupled-cluster-level reference calculations to within 0.05 kcal/mol. Our calculations differ from those of Nagy et al., however, in that we used F12b as the F12 method for CCSD (instead of F12*/F12c), and performed a [D/T] CBS extrapolation on the triples from the CCSD(T)-F12/haV[D/T]Z calculations (rather than using the [T/Q] extrapolated triples from a pair of CCSD(T)/haV[T/Q]Z calculations); we thus refer to our calculations as "electrum standard," as electrum is a mixture of gold and silver. The interaction energy is thus the sum of DF-HF+CABS/aVQZ-F12 with half-CP, DF-MP2-F12/aV[T/Q]Z-F12 with full-CP, and CCSD(T)-F12b/haV[D/T]Z – DF-MP2-F12/haV[D/T]Z with full-CP. (For some aromatic-containing systems (e.g., benzene dimers), we had to delete the diffuse *f* functions on the carbons in the haVQZ-F12 basis to avoid a basis-set linear-dependence issue (also described in Ref. 64). For a few benzene-containing systems (dimers with itself, water, methanol, phenol), we used full-CP HF/aVQZ-F12 instead of half-CP to avoid basis-set issues that arose only for the isolated benzene monomers. For phenol dimers and phenol-benzene dimers, we also had to delete the diffuse *f* functions on the carbons in the haVTZ-F12 basis to avoid CCSD convergence issues.) We used extrapolation coefficients as suggested previously,[73] 4.6324 for [T/Q] and 3.1458 for [D/T], in a two-point scheme. Calculations under these conditions match the 14K-GOLD standard of Nagy et al. to within 0.03 kcal/mol (worst 0.03 kcal/mol, root-mean-square deviation 0.01 kcal/mol, bias −0.005 kcal/mol) on the S66 benchmark set.

### *B. Results*

Our benchmarking results are summarized in Tables X.I–X.III below. The reference for dimers is always the electrum-standard composite interaction energy, described above. The reference for microsolvation nonadditivity is MP2 with a mixed basis set, as noted in section X.B.3. The labels and acronyms used are:

**System** – The chemical system in question. Neat systems are a single name (e.g., “water”), while mixture systems have a “/” separating the two components (e.g., “benzene/water”).

**Basis** – the basis of the CC-level calculation performed.

**wRMSE** – (weighted) root-mean-square error

**wMSE** – (weighted) mean signed error (method minus reference, so that positive values indicate under-binding)

**Opt Dimer Scans** – Weighted interaction energies of scans generated from geometry-optimized dimers.

**Microsolvation Dimer Sums** – Sum of solute-solvent dimer interaction energies present in a microsolvation cluster (i.e., the two-body part of the microsolvation energy). Only the first 10 microsolvation geometries for each system were used in benchmarking.

1. *Comparison to CCSD(T)/aVTZ*

For water only, our models were fit to data from a composite scheme utilizing full-CP CCSD(T)/aVTZ calculations.

**Table X.I.** Benchmark statistics (in kcal/mol) of the theory used to generate our training data for (neat) water only: CCSD(T)/aVTZ + DF-MP2/aV[T/Q]Z + DF-HF/aVQZ with full-CP corrections, compared to an electrum-standard reference, as described in section X.A.

| **System** | **Basis** | **Opt Dimer Scans** | | **Microsolvation Dimer Sums** | |
|---|---|---|---|---|---|
| | | **wRMSE** | **wMSE** | **RMSE** | **MSE** |
| Water | aVTZ | 0.00 | 0.00 | 0.03 | 0.03 |

### 2. *Comparison to PNO-LCCSD(T*)-F12b/aVXZ-F12*

For all systems except water, a composite scheme utilizing PNO-LCCSD(T*)-F12b/aVXZ-F12 was used. Separate CP schemes were chosen for the CCSD and (T) components, on a per-system basis. (Our choice of CP correction scheme changed over time in an attempt to follow best practices for gold-standard calculations in the most recent literature. We always used either half-CP or full-CP, and the difference between the two is less than the error of our force field fits.) In addition, the PNO code within Molpro changed between versions 2019.2.3 and 2020.1.2 of the software, including a modification to the cutoff values set by "DOMOPT=TIGHT," which led to a less strict energy cutoff for inclusion of orbital pairs.

**Table X.II.** Benchmark statistics (in kcal/mol) of the theory we used to generate our training data for all systems except neat water: PNO-LCCSD(T*)-F12b/aVXZ-F12 + DF-MP2/aV[T/Q]Z + DF-HF/aVQZ, compared to an electrum-standard reference as described in section X.A. The version of Molpro used to generate the data is included here because the defaults in the PNO code were changed in the 2020.1 release of Molpro,[57] altering how well the PNO energies matched their canonical references.

| **System** | **Basis** | **Molpro Version** | **Dimer Scans** | | **Microsolvation Dimer Sums** | |
|---|---|---|---|---|---|---|
| | | | **wRMSE** | **wMSE** | **RMSE** | **MSE** |
| Benzene | aVDZ-F12 | 2019.2.3 | 0.04 | 0.04 | 0.11 | 0.11 |
| Benzene/Water | aVTZ-F12 | 2019.2.3 | 0.03 | 0.03 | 0.05 | −0.02 |
| Methanol | aVQZ-F12 | 2019.2.3 | 0.02 | 0.01 | 0.03 | 0.00 |
| Methanol/Water | aVQZ-F12 | 2019.2.3 | 0.03 | 0.02 | 0.06 | 0.00 |
| Methanol/Benzene | aVQZ-F12 | 2019.2.3 | 0.01 | −0.01 | 0.05 | 0.03 |
| Ethane | aVQZ-F12 | 2020.1.2 | 0.01 | −0.01 | 0.01 | 0.01 |
| Ethane/Water | aVQZ-F12 | 2020.1.2 | 0.01 | 0.01 | 0.07 | 0.06 |
| Ethylene | aVQZ-F12 | 2020.1.2 | 0.01 | 0.00 | 0.03 | 0.01 |
| Ethylene/Water | aVQZ-F12 | 2020.1.2 | 0.02 | 0.02 | 0.19 | 0.15 |
| Acetone | aVTZ-F12 | 2020.1.2 | 0.06 | 0.04 | 0.19 | 0.19 |
| Acetone/Water | aVQZ-F12 | 2020.1.2 | 0.04 | 0.04 | 0.17 | 0.16 |
| Acetone/Benzene | aVTZ-F12 | 2020.1.2 | 0.03 | 0.03 | 0.25 | 0.25 |
| Dimethylamine | aVTZ-F12 | 2020.2.0 | 0.04 | 0.04 | 0.13 | 0.12 |
| Dimethylamine/Water | aVQZ-F12 | 2020.2.1 | 0.02 | 0.02 | 0.16 | 0.16 |
| Dimethylamine/Benzene | aVDZ-F12 | 2020.2.1 | 0.09 | 0.09 | 0.26 | 0.25 |
| Dimethyl sulfide | aVTZ-F12 | 2020.2.0 | 0.02 | 0.02 | 0.13 | 0.12 |
| Dimethyl sulfide/Water | aVQZ-F12 | 2020.2.1 | 0.02 | 0.02 | 0.11 | 0.10 |
| Pyrrole | aVTZ-F12 | 2020.2.0 | 0.07 | 0.06 | 0.34 | 0.34 |
| Pyrrole/Water | aVQZ-F12 | 2020.2.1 | 0.03 | 0.03 | 0.21 | 0.20 |
| Pyrrole/Benzene | aVDZ-F12 | 2020.2.1 | 0.13 | 0.12 | 0.41 | 0.40 |
| NMA | aVTZ-F12 | 2020.2.1 | 0.05 | 0.05 | 0.14 | 0.14 |
| NMA/Water | aVQZ-F12 | 2020.2.1 | 0.04 | 0.04 | 0.21 | 0.19 |
| NMA/Benzene | aVDZ-F12 | 2021.2.1 | 0.03 | 0.03 | 0.11 | 0.10 |
| Phenol | aVDZ-F12 | 2020.2.1 | 0.18 | 0.17 | 0.64 | 0.63 |
| Phenol/Water | aVTZ-F12 | 2020.2.1 | 0.06 | 0.05 | 0.38 | 0.34 |
| Phenol/Benzene | aVDZ-F12 | 2020.2.1 | 0.12 | 0.12 | 0.49 | 0.48 |
| Imidazole | aVTZ-F12 | 2020.2.1 | 0.09 | 0.08 | 0.36 | 0.36 |
| Imidazole/Water | aVQZ-F12 | 2020.2.1 | 0.04 | 0.04 | 0.23 | 0.21 |

### 3. *Comparison to ωB97M-V/def2-TZVPPD*

The ωB97 family of functionals remains a popular choice for large-scale datasets designed for training machine-learning force fields and structure-property prediction methods.[29–32,74,75] ωB97M-V performs especially well on non-covalent interactions,[76] and when paired with the def2-TZVPPD basis set, offers a very affordable compromise between cost and accuracy. Here, we used ωB97M-V/def2-TZVPPD with full-CP corrections, which we found to produce the best match to our electrum-standard calculations. Calculations were performed using the Psi4 software package.

For "microsolvation nonadditivity," the reference used was the DF-MP2/aV[D/T]Z composite calculations used to fit our force fields. The DFT values are overly attractive in nonadditivity compared to MP2, a finding that cannot be attributed to the known lack of MBD effects in MP2, because these effects are generally repulsive in nature. Interestingly, the overly attractive DFT nonadditivities mostly cancel out the overly repulsive dimer interactions; the total microsolvation energies are thus (fortuitously) accurate through a cancellation of errors. We also note that the errors in optimized dimer scans and the errors over dimers in microsolvation systems are sometimes opposite in sign, suggesting that results from optimized dimers (e.g., from S66 or S66x8) may not be the only type of dimer calculations one should use when assessing QM accuracy.

**Table X.III.** Benchmark statistics (in kcal/mol) of ωB97M-V/def2-TZVPPD compared to, for dimers, an electrum-standard reference, as described in section X.A. For microsolvation nonadditivity, the reference is the DF-MP2/aV[D/T]Z composite calculations used to fit our force fields.

| System | Dimer Scans | | Microsolvation Dimer Sums | | Microsolvation Nonadditivity | |
|---|---|---|---|---|---|---|
| | wRMSE | wMSE | RMSE | MSE | RMSE | MSE |
| Water | 0.04 | 0.04 | 0.21 | 0.21 | 0.43 | −0.41 |
| Benzene | 0.10 | −0.04 | 1.13 | 1.11 | 0.65 | −0.62 |
| Benzene/Water | 0.07 | −0.05 | 0.70 | 0.57 | 0.72 | −0.69 |
| Methanol | 0.27 | 0.24 | 0.78 | 0.74 | 0.72 | −0.71 |
| Methanol/Water | 0.10 | 0.09 | 0.72 | 0.64 | 0.81 | −0.77 |
| Methanol/Benzene | 0.12 | 0.05 | 1.05 | 0.98 | 0.87 | −0.79 |
| Ethane | 0.12 | 0.12 | 0.80 | 0.79 | 0.71 | −0.70 |
| Ethane/Water | 0.13 | 0.13 | 0.71 | 0.65 | 0.76 | −0.71 |
| Ethylene | 0.11 | 0.09 | 0.75 | 0.74 | 0.72 | −0.69 |
| Ethylene/Water | 0.07 | −0.07 | 0.53 | 0.42 | 0.66 | −0.62 |
| Acetone | 0.16 | 0.09 | 0.66 | 0.62 | 0.65 | −0.63 |
| Acetone/Water | 0.05 | 0.01 | 0.74 | 0.59 | 1.12 | −1.02 |
| Acetone/Benzene | 0.13 | −0.11 | 0.92 | 0.90 | 0.90 | −0.88 |
| Dimethylamine | 0.16 | 0.13 | 1.02 | 0.99 | 0.75 | −0.72 |
| Dimethylamine/Water | 0.19 | 0.17 | 1.00 | 0.92 | 0.87 | −0.81 |
| Dimethylamine/Benzene | 0.12 | −0.11 | 1.21 | 1.19 | 1.04 | −0.98 |
| Dimethyl sulfide | 0.22 | 0.20 | 1.56 | 1.53 | 1.20 | −1.18 |
| Dimethyl sulfide/Water | 0.10 | 0.09 | 0.91 | 0.80 | 1.02 | −0.93 |
| Pyrrole | 0.25 | 0.20 | 1.59 | 1.56 | 1.41 | −1.39 |
| Pyrrole/Water | 0.14 | 0.06 | 0.89 | 0.78 | 1.22 | −1.13 |
| Pyrrole/Benzene | 0.09 | 0.07 | 1.57 | 1.54 | 1.29 | −1.25 |
| NMA | 0.09 | 0.04 | 0.99 | 0.94 | 1.03 | −1.01 |
| NMA/Water | 0.13 | 0.09 | 0.74 | 0.65 | 1.10 | −1.00 |
| NMA/Benzene | 0.12 | −0.11 | 1.07 | 1.06 | 1.27 | −1.25 |
| Phenol | 0.22 | −0.17 | 1.24 | 1.23 | 1.76 | −1.74 |
| Phenol/Water | 0.18 | 0.09 | 1.09 | 0.96 | 1.40 | −1.30 |
| Phenol/Benzene | 0.12 | 0.09 | 1.63 | 1.60 | 1.61 | −1.58 |
| Imidazole | 0.29 | 0.15 | 1.94 | 1.92 | 1.44 | −1.43 |
| Imidazole/Water | 0.18 | 0.15 | 0.93 | 0.85 | 1.14 | −1.01 |